\documentclass{aa}  

\usepackage{graphicx}
\usepackage{subcaption}
\usepackage{txfonts}
\usepackage{makecell}
\usepackage{hyperref}
\usepackage{cleveref}

\crefname{equation}{Eq.}{Eqs.}
\Crefname{equation}{Equation}{Equations}
\crefname{figure}{Fig.}{Figs.}
\Crefname{figure}{Figure}{Figures}
\crefname{table}{Table}{Tables}
\Crefname{table}{Table}{Tables}
\crefname{chapter}{Chapter}{Chapters}
\Crefname{chapter}{Chapter}{Chapters}
\crefname{section}{Section}{Sections}
\Crefname{section}{Section}{Sections}
\crefname{appendix}{Appendix}{Appendices}
\Crefname{appendix}{Appendix}{Appendices}

\newcommand{\msol}{\,M$_{\odot}$ }
\newcommand{\fcbm}{f_{\rm CBM}}
\newcommand{\rcc}{r_{\rm cc}}
\newcommand{\rcbm}{r_{\rm CBM}}
\newcommand{\lb}{\left(}
\newcommand{\rb}{\right)}
\newcommand{\n}{\nabla_{\text{T}}}
\newcommand{\nad}{\nabla_{\text{ad}}}
\newcommand{\nrad}{\nabla_{\text{rad}}}
\newcommand{\Dcbm}{D_{\text{CBM}}}
\newcommand{\Dmix}{D_{\text{mix}}}
\newcommand{\Hp}[1][\rm cc]{H_{\rm p, {#1}}}

\makeatletter
\renewcommand*\aa@pageof{, page \thepage{} of \pageref*{LastPage}}
\makeatother

\begin{document} 

   \title{Probing the shape of the mixing profile and of the  thermal structure at the convective core boundary through asteroseismology}

   \subtitle{}

   \author{M. Michielsen
          \inst{1}
          \and M.G. Pedersen
          \inst{1}
          \and K. C. Augustson
          \inst{2}      
          \and S. Mathis
          \inst{2}
          \and C. Aerts
          \inst{1,3}          
          }

   \institute{\inst{1}Institute of Astronomy, KU Leuven, Celestijnenlaan 200D, B-3001 Leuven, Belgium\\
             \inst{2} AIM, CEA, CNRS, UniversitÃ© Paris-Saclay, UniversitÃ© Paris Diderot, Sarbonne Paris CitÃ©, F-91191 Gif-sur-Yvette Cedex, France \\
             \inst{3}Department of Astrophysics, IMAPP, Radboud University Nijmegen, P.O. Box 9010, 6500 GL Nijmegen, The Netherlands \\
             \email{mathias.michielsen@kuleuven.be}}

   \date{Received 23 April 2019; accepted 11 June 2019}

 
  \abstract
   {}
   { We investigate from a theoretical perspective if space asteroseismology can
     be used to distinguish between different thermal structures and shapes of
     the near-core mixing profiles for different types of coherent oscillation
     modes in massive stars with convective cores; we also examine whether this capacity depends
     on the evolutionary stage of the models along the main sequence.}
 {We computed 1D stellar structure and evolution models for four different
   prescriptions of the mixing and temperature gradient in the near-core
   region. We investigated their effect on the frequencies of dipole prograde gravity modes 
in slowly pulsating B stars and in  $\beta$ Cep stars as well as pressure modes in $\beta$ Cep stars.}
   {A comparison between the mode frequencies of the different models at various
     stages during the main sequence evolution reveals that they are
     more sensitive to a change in temperature gradient than to the exact shape
     of the mixing profile in the near-core region. Depending on the
     duration of the observed light curve, we can distinguish between either
     just the temperature gradient, or also between the shapes of the mixing
     coefficient. The relative frequency differences are in general larger for
     more evolved models and are largest for the higher frequency pressure
     modes in $\beta$ Cep stars.}
   {In order to unravel the core boundary mixing and thermal
       structure of the near-core region, we must have asteroseismic
       masses and radii with $\sim 1\%$ relative precision for hundreds of stars.} 

   \keywords{Asteroseismology -- convection --
                stars: oscillations (including pulsations) -- stars: interiors
                -- techniques: photometric }
  \titlerunning{Mixing profile and thermal structure at the 
convective core boundary}
   \maketitle
%

\section{Introduction} 

There are some hurdles left to overcome in stellar evolution models before they meet the precision of observed diagnostics from $\mu$mag space photometry and spectroscopy. One of these is the mass discrepancy between theoretical and model-independent dynamical masses in binary systems. Masses derived from the orbital solution of the system are found to be lower than the masses required in theoretical models to fit the location in the Hertzsprung-Russell diagram by means of isochrones. A prominent example of this is the binary system V380 Cyg \citep{2000ApJ...544..409G}.  After a detailed analysis, \citet{2014MNRAS.438.3093T} found the masses derived from high-precision \textit{Kepler} space photometry to be higher than the dynamical masses as well, and concluded that current single-star evolutionary models are not suitable to reproduce the observed properties of the binary. More specifically, these authors concluded that a large amount of core mass is lacking.  Similar cases in which a large amount of overshooting was needed to reconcile theoretical and dynamical masses include the $\theta$ Ophiuchi system \citep{2007MNRAS.381.1482B} and the V578 Mon system \citep{2014AJ....148...39G}. There are numerous other systems in which a mass discrepancy has been observed, for example LMC 172231, ST2-28, LMC 169782, LMC 171520, and [P93] 921 \citep{2012ApJ...748...96M,2014ApJ...789..139M}.

This near-core mixing, required to get more mass into the convective core fixed by the Ledoux criterion, is likely a combination of several physical processes such as convective overshooting, internal gravity waves, shear instabilities due to rotation, mean meridional flows, and magnetism \citep[see e.g. the detailed discussion in][]{2012A&A...539A..90N}. From a theoretical point of view, a vast number of free parameters is used for the overall computation of the mixing profile, $\Dmix(r)$, throughout the radiatively stratified envelope of the star. As discussed in for example \citet{2000A&A...361..101M} and \citet{2013A&A...558A.103G}, there is no reason from the modelling perspective to prefer one above many other descriptions of various ingredients of $\Dmix(r)$. Therefore, we take an empirical approach in this study and consider the simplest case in which the rotation of the star is slow enough to work with non-rotating 1D equilibrium models. However the oscillations are treated taking into account the Coriolis acceleration under the assumption of rigid rotation. The latter is justified given the very low level of differential rotation in asteroseismic inferences obtained for slowly pulsating B stars (SPB) and $\beta$ Cep stars \citep{2018A&A...618A..24V}. The approach to compute the oscillation frequencies from perturbing non-rotating equilibrium models is valid as long as the rotation of the star remains below $\sim\! 50\%$ of the critical Roche rotation, such that the centrifugal acceleration, which is responsible for the flattening of the star, can be ignored \citep[e.g.][]{2017MNRAS.465.2294O}. The treatment of the Coriolis acceleration is done using the traditional approximation of  rotation (TAR) \citep[e.g.][]{1960PhFl....3..421E,1989nos..book.....U,1996ApJ...460..827B} for the gravity (g) modes. The TAR neglects the horizontal component of the angular velocity vector in the Coriolis acceleration, causing the perturbed equations of stellar structure to become separable in radial and tangential coordinates. This set of equations is known as Laplace's tidal equations \citep{2003MNRAS.340.1020T}. Its solutions provide eigenvectors that are an excellent approximation for the g modes of SPBs, whose tangential component is completely dominant over the radial component. Since the TAR is only meaningful for g modes and the rotational frequencies are small compared to the frequencies of pressure (p)
modes, a first-order perturbative approach known as Ledoux splitting is adequate and used in the treatment of the p modes \citep{1958HDP....51..353L}. As extensively discussed and illustrated by \citet[][Figs\,2 and 3]{2019A&A...624A..75A}, this approach is appropriate for p and g modes detected in \textit{Kepler} data of SPBs and $\beta\,$Cep stars.

Various types of parameterisations for boundary mixing profiles can be used based on numerical simulations as well (see \citet{2015A&A...580A..61V} for an in depth discussion).  Indeed, more and more 3D global nonlinear numerical simulations of the (magneto-)hydrodynamics of the convective cores of early-type stars are now computed \citep{2004ApJ...601..512B,2005ApJ...629..461B,2013ApJ...772...21R,2015ApJ...815L..30R,2016ApJ...829...92A,Edelmann_2019} that provide us information and predictions on convective penetration or overshoot in their surrounding, stably stratified radiative envelopes.  We can make different assumptions regarding these processes, which we divide in two categories. The first concerns the shape of the mixing profile $\Dmix(r)$, which determines the extent and efficiency of the near-core mixing.  The second are the assumptions regarding the thermal structure in the near-core region. Assuming this to be radiative or adiabatic entails a diffusive or convective mixing process, respectively.  The local P\'eclet number, which compares to thermal diffusivity, provides a means of assessing which of the mixing models is most appropriate in this region \citep[as in][]{2015A&A...580A..61V}.

The intention of this paper is to compare various prescriptions for near-core mixing. Specifically, we wish to see if they are seismically distinguishable in main sequence stars with a convective core and a radiative envelope under the assumption that the stellar mass is known. We study the effect of a change in the shape of the mixing profile near the convective boundary. Furthermore, we also investigate the effect of changing the temperature gradient in the region with core boundary mixing. We study whether or not it is possible to distinguish between these near-core mixing prescriptions, and if the ability to do so is dependent on the evolutionary stage during the main sequence and on the type of pulsations. Therefore, B stars undergoing coherent pressure (p-) or gravity (g-) mode pulsations \citep[e.g.][]{2010aste.book.....A} are considered since the effect of missing core mass is larger the higher the birth mass of the star. In practice, we try to answer the question if observed coherent non-radial oscillation modes detected in \textit{Kepler} photometry are able to probe the shape of the near-core mixing profile and the thermal structure of this region for stars with a fully mixed convective core that retreats as the star evolves. As a first step, we consider stars rotating sufficiently slowly so that deformation may be ignored in order to evaluate the efficacy of these oscillatory modes in probing the near-core mixing mechanisms.

\section{Near-core mixing prescriptions developed in \texttt{MESA}}\label{sect:mixing_prescriptions}

\begin{figure*}[ht]
    \centering 
    \begin{subfigure}{0.29\hsize}
        \includegraphics[width=\hsize]{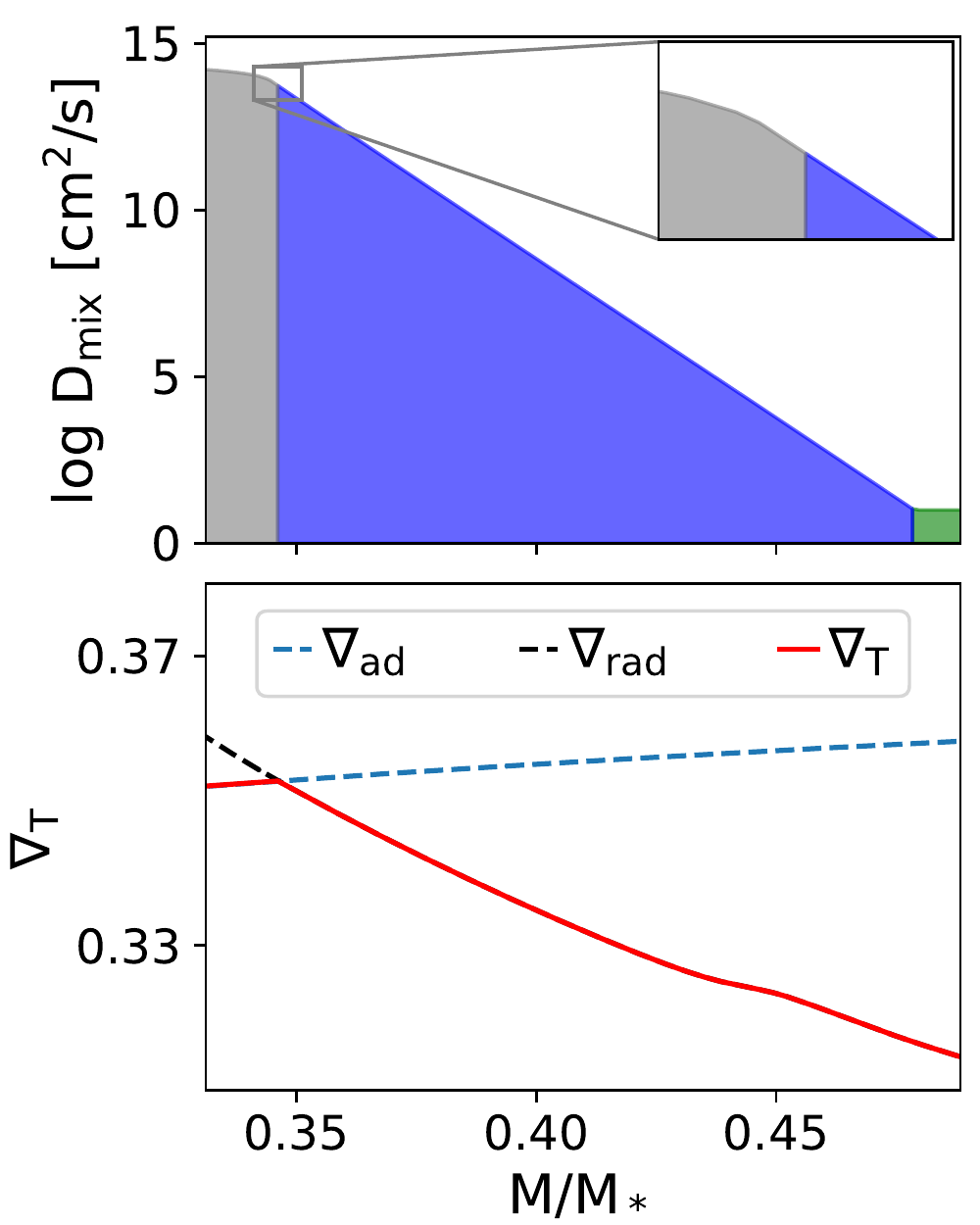}
        \caption{\\ Exponential overshoot}
        \label{fig:profile_exponential}
    \end{subfigure}
    \begin{subfigure}{0.232\hsize}
        \includegraphics[width=\hsize]{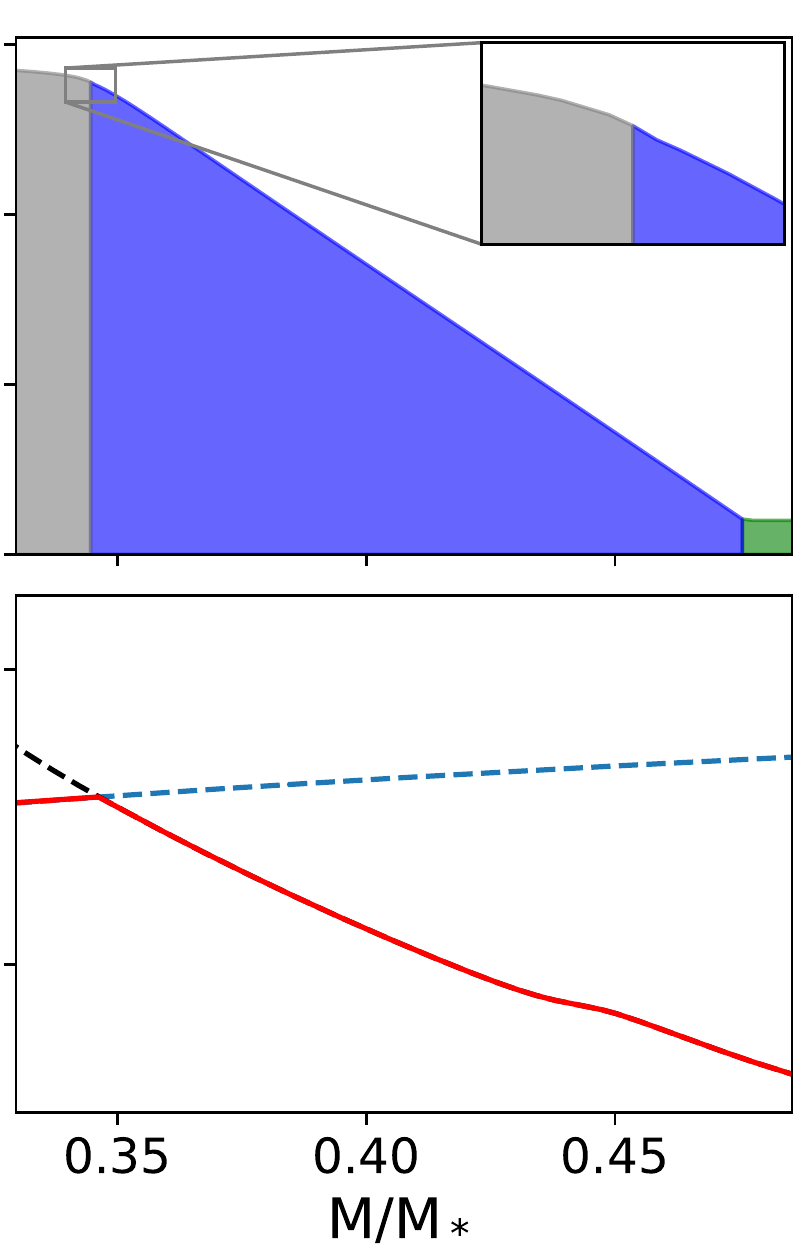}
        \caption{\\ Diffusive Gumbel profile}
        \label{fig:profile_gumbel}
    \end{subfigure}
    \begin{subfigure}{0.232\hsize}
        \includegraphics[width=\hsize]{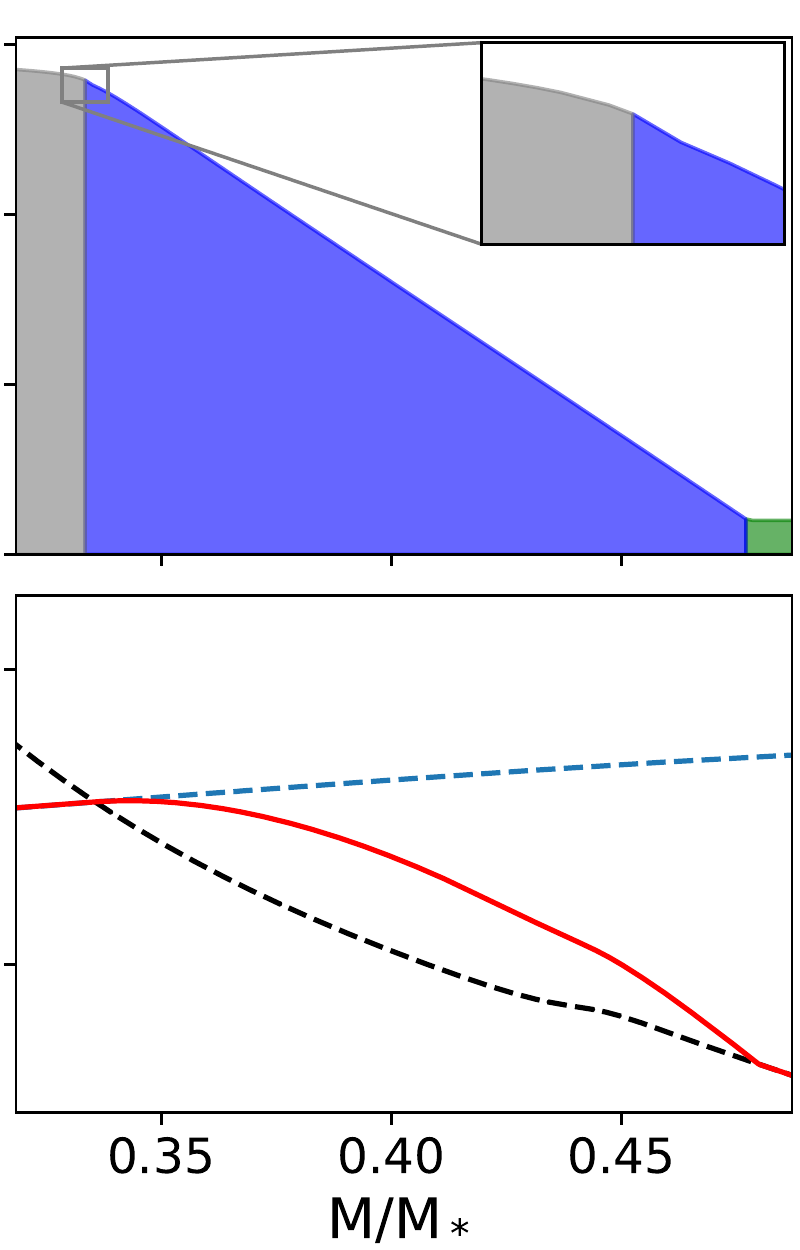}
        \caption{\\ Convective Gumbel profile}
        \label{fig:profile_gumbel_gradT}
    \end{subfigure}
    \begin{subfigure}{0.232\hsize}
        \includegraphics[width=\hsize]{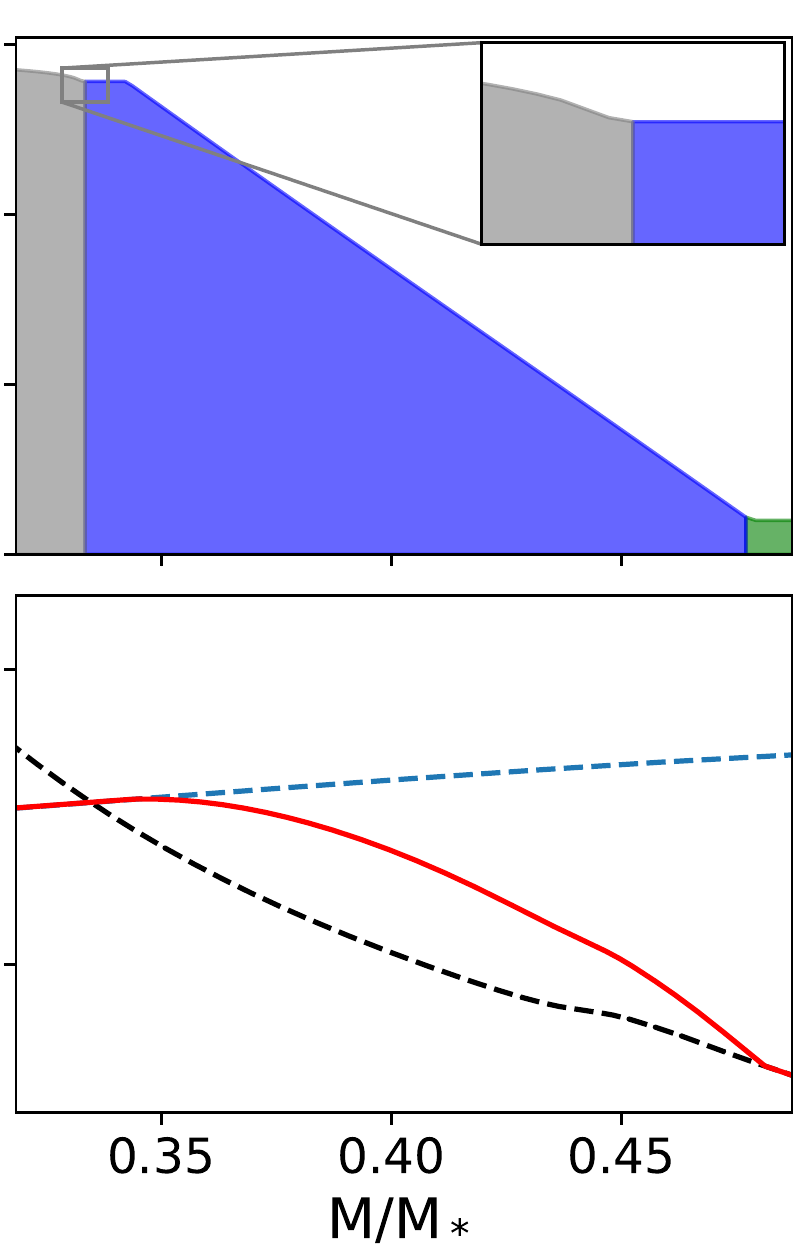}
        \caption{\\ Extended convective penetration}
        \label{fig:profile_extended_core}
    \end{subfigure}    
    \caption{Top panels show different shapes of near-core mixing profiles for a star with a mass of 12\msol near the ZAMS. The convective core is indicated in grey, the near-core mixing region in blue, and constant diffusive mixing in the outer radiative envelope in green. The bottom panels show the temperature gradients in the same region as the top panels. In each figure the insets are zoomed in on the same region of m/M$_{*}$ and D$_{\text{mix}}$ to show the differences at the edge of the core mixing region more clearly.  }
    \label{fig:profiles}
\end{figure*}

In this work, the code \texttt{MESA} is used, which is diffusive by construction. We use version 10398 of the code, as described in \citet{2018ApJS..234...34P}. A few near-core mixing prescriptions are standardly implemented in \texttt{MESA}. These consist of a step-like overshoot profile, an exponentially decaying profile, and an extended exponential profile, all of which adopt the radiative temperature gradient outside the convective core. The ability of g modes to distinguish between those profiles has been investigated by \citet{2018A&A...614A.128P}. Next to the exponentially decaying profile as implemented in \texttt{MESA}, we present three newly implemented near-core mixing prescriptions, which are illustrated in \cref{fig:profiles} and discussed in the following subsections.

\subsection{Diffusive exponential overshooting}\label{sect:diff_exp}

Exponential overshoot is one of the options available in \texttt{MESA}. It was motivated by 2D hydrodynamical simulations of white dwarfs and A-type stars \citep{1996A&A...313..497F}, and AGB stars \citep{2000A&A...360..952H}. It assumes a decrease in efficiency of mixing further away from the convective core boundary $\rcc$, starting with the mixing coefficient $D_0$ at the inner side of the convective core boundary resulting from mixing length theory \citep[MLT;][]{1958ZA.....46..108B}, and decreasing further outward

\begin{equation} \label{eq:exp_mix}
    \Dcbm(r) = D_0 \exp{\lb \frac{-2 (r-r_0)}{\fcbm \Hp} \rb}.
\end{equation}
In this equation, $\rcc$ is defined as the core radius resulting from the Schwarzschild criterion of convection. For the computation of the models, we rely on the Ledoux criterion, which takes into account the influence of the chemical gradient left behind by the receding convective core characteristic of the B-type stars considered in this study. The extent of this exponentially decaying region is determined by the pressure scale height at the edge of the convective core $\Hp$ and a free parameter $\fcbm$. Although $\fcbm$ is usually written as $f_{\rm ov}$ and associated with overshooting, we use this notation to parameterise the extent of near-core mixing region for different prescriptions, where we adapt the subscript CBM to denote core boundary mixing. The transition between core mixing and overshoot is made from inside the convective core at $r_0 = \rcc - f_0 \Hp$; $D_0$ is the value of the mixing coefficient at this radius, and is hence influenced by the choice of $f_0$, which is typically set smaller than $\fcbm$. \Cref{fig:profile_exponential} depicts the shape of the near-core mixing profile in the case of this diffusive exponential treatment of core overshooting.
This approach assumes that the thermal structure in the overshoot zone ($r>\rcc$) is $\n = \nrad$, and $\n = \nad$ in the convective core ($r<\rcc$), i.e. 

\begin{align}
    \nrad =& \left( \frac{\partial \ln T}{\partial \ln p} \right)_{\text{rad}} = \frac{3}{16 \pi acG} \frac{\kappa l P}{m(r) T^4}, \\
    \nad  =& \left( \frac{\partial \ln T}{\partial \ln p} \right)_{\text{ad}},
\end{align}
where $a$ is the radiation constant, $c$ the speed of light, $G$ the
gravitational constant, $\kappa$ the local Rosseland opacity,
$l$ the local luminosity, $P$ the local pressure, $T$ the local temperature, and
$m$ the mass inside a spherical shell with radius $r$ measured from the stellar
centre. The overshoot region expands to $\rcbm$, which is the position in the
outer envelope where a constant level of $\Dmix$ takes over (green area in
\cref{fig:profiles}). 

\subsection{Diffusive Gumbel overshooting} \label{sect:gumbel_profile} Another
way to parameterise diffusive mixing processes is to use a profile of the mixing
coefficient based upon the Gumbel distribution. \citet{2017A&A...604A.125P}
proposed this functional form instead of a regular exponential decay based upon
2D hydrodynamical simulations of the young Sun.  This diffusion coefficient is
derived to describe flows in a large P\'eclet number regime (Pe$\gg$1), which is
characteristic of stellar interiors. This is in contrast to the exponentially decaying diffusion coefficient by \citet{1996A&A...313..497F},
which is derived for flows in the low P\'eclet number regime (Pe$\ll$1) such as
those in stellar envelopes.  In our models, the P\'eclet number goes from the
high regime at the core boundary to the low regime at the outer edge of the
boundary mixing region. Therefore, to apply each prescription in their valid
regime, we should switch approximately halfway through the near-core mixing
region from the Gumbel profile to the exponential prescription. However, we
first treat these as two separate functional forms of the diffusion coefficient
to test the influence of the latter on the mode frequencies.

Using equation (70) from \citet{2019ApJ...874...83A}, but simplifying for the non-rotating case, yields 

\begin{equation}
    \Dcbm(r) = D_0 \lb 1-\exp \lb -\exp \lb \frac{r-\rcc}{\lambda L_P}+\frac{\mu}{\lambda} \rb \rb \rb,
\end{equation}
where $\mu$ and $\lambda$ are variable parameters and $L_P$ is the convective penetration depth, which depends upon rotation in \citet{2019ApJ...874...83A}. However, since only the non-rotating case is treated, $L_P$ depends upon the pressure scale height and the values of many physical quantities as shown in equations (67) and (68) of \citet{2019ApJ...874...83A}.
In order to compare with \cref{eq:exp_mix},
$\lambda L_P$ and $\mu L_P$ are replaced by $\fcbm \Hp$ and $f_0 \Hp$, respectively.
The parameters $\mu$ and $\lambda$ are constrained by the simulations of \citet{2017A&A...604A.125P}. So while they are free to some degree, some care should be taken when directly mapping from $\lambda L_P$ to $\fcbm \Hp$. Specifically, $\fcbm$ can directly be related to those terms in equations (67) and (68) in \citet{2019ApJ...874...83A}.
Incorporating these substitutes and normalising the expression to yield $\Dcbm = D_0$ at $r_0$, results in

\begin{equation}
    \Dcbm(r) = D_0 \lb \frac{1- \exp \lb -\exp \lb -\left| \frac{r-r_0}{\fcbm \Hp}\right|  \rb \rb}{1-\exp (-1)} \rb.
\end{equation}
The absolute value is taken to ensure this expression can be used for models with convective cores, where the overshoot is directed towards the surface, as well as for models with radiative cores and convective envelopes, in which the overshoot is directed towards the stellar core.  
This kind of mixing profile is illustrated in \cref{fig:profile_gumbel}, where the thermal structure is again assumed to be $\n = \nrad$ outside the convective core ($r>\rcc$).

Comparing this profile in \cref{fig:profile_gumbel} to the diffusive exponential profile in \cref{fig:profile_exponential}, it is difficult to spot the difference in mixing coefficient on this scale. The
difference only becomes apparent when looking at the insets, which are zoomed in on the exact same region near the convective core boundary. When going from the convective core outwards, the mixing coefficient initially decays slower in the Gumbel profile. However, the decay becomes faster than in the exponential case when moving further outwards, making both profiles extend to the same radius where the constant outer envelope mixing takes over.

\subsection{Convective Gumbel penetration}
Overshooting material can influence the entropy stratification in cases in which the convective boundary is located deep in the stellar interior, as shown by \citet{1991A&A...252..179Z}. This ``convective penetration'' entails a nearly adiabatic temperature gradient in the near-core mixing region. The P\'eclet number near the edge of the convective core is much larger than unity, but drops multiple orders of magnitude over the extent of the near-core mixing region, becoming much smaller than unity at the outer edge of this region. We therefore let the temperature gradient make a gradual transition from fully adiabatic $\n = \nad$ in the convective core to fully radiative $\n = \nrad$ in the radiative outer layers.
Using the same prescription for the diffusion coefficient as in the previous section, but adjusting the temperature gradient in the near-core mixing region to make this gradual transition gives a profile as shown in \cref{fig:profile_gumbel_gradT}. The gradient in this transition region is calculated as

\begin{equation}
    \n = g\nad + (1-g)\nrad \quad \text{for} \quad  \rcc < r < \rcbm,
\end{equation}
where the factor $g$ is given by

\begin{equation}\label{eq:fcc}
    g = \frac{q(\rcbm) - q(r)}{ q(\rcbm)-q(\rcc)  }.
\end{equation}
In this equation, $q$ denotes the relative mass coordinate.

\subsection{Extended convective penetration}
As explained in the introduction, modelling of stars in binary systems requires an increased amount of mass in the convective core to make the evolutionary masses match the dynamical masses. This can be addressed by enlarging the core by means of convective penetration.
The latter is included in some stellar evolution codes\footnote{For example CL\'ES \citep{2008Ap&SS.316...83S} and \texttt{GENEVA} \citep{2008Ap&SS.316...43E}.} by introducing a step-like overshoot function, $\Dcbm = D_0$, over a distance of $\alpha \Hp$, where $\alpha$ is a free parameter. The temperature gradient in this region is fully adiabatic $\n = \nad$.
This prescription entails that at $r=r_0+\alpha \Hp$, the temperature gradient discontinuously switches from fully adiabatic to fully radiative, and the mixing coefficient drops from D$_{0}$ to the small amount of diffusive mixing present in the radiative envelope.

It was shown by \citet{2015A&A...580A..27M,2016ApJ...823..130M} that an exponentially decaying overshoot performs better asteroseismically than a step-like overshoot function to describe the seismic data in two SPB stars when the radiative temperature gradient is taken in the overshoot zone.
Therefore, instead of taking the step-like mixing coefficient of the classical convective penetration, an exponential decay as discussed in \cref{sect:diff_exp} is introduced on top of the traditional convective penetrative region. 
Additionally, to avoid the discontinuity in the temperature gradient entailed by the classical convective penetration, the temperature gradient in the exponentially decaying region is set to gradually switch from fully adiabatic in the penetrative region towards fully radiative in the envelope, as illustrated in \cref{fig:profile_extended_core}.
The mixing coefficient is described by

\begin{align}
    \Dcbm(r) &= D_0 && \text{for} \quad  r_0<r<r_{\rm cp} ,\\
    \Dcbm(r) &= D_0 \exp{\lb \frac{-2 (r-r_{\rm cp})}{\fcbm \Hp[cp] }\rb}  &&\text{for} \quad  r_{\rm cp} < r < \rcbm, \label{eq:ext_exp_mix}
\end{align}
along with the following temperature gradient:
\begin{align}
    \n &= \nad   &&\text{for} \quad   r<r_{\rm cp} ,\\
    \n &= h\nad + (1-h)\nrad  &&\text{for} \quad  r_{\rm cp} < r < \rcbm ,\\
    \n &= \nrad &&\text{for} \quad \rcbm < r,
\end{align}
where $r_{\rm cp} = r_0+\alpha \Hp$ is the edge of the traditional step-like penetration region and factor $h$ is similar to $g$ in \cref{eq:fcc}
\begin{equation}\label{eq:fcp}
    h = \frac{q(\rcbm) - q(r)}{ q(\rcbm)-q(r_{\rm cp})}.
\end{equation}
We note that since the exponential decay starts at the edge of the convective penetrative region, $\Hp[cp]$ is used in \cref{eq:ext_exp_mix} instead of $\Hp$, denoting that the pressure scale height at the edge of the penetration region is used instead of at the edge of the convective core.
Compared to the prescriptions previously explained, the extended convective penetration has an extra free parameter $\alpha$. In addition to $\fcbm$, which governs the exponential decay, $\alpha$ is introduced to determine the extent of the step-like mixing region, resulting in an increase of the convective core mass. 

Henceforth, the diffusive exponential profile and diffusive Gumbel profile are referred to as diffusive profiles, while the convective Gumbel profile and extended convective penetration are referred to as convective profiles for simplicity.

\section{Computational set-up}
\subsection{\texttt{MESA} models}
To test the effect of the different near-core mixing prescriptions and the thermal structure on the pulsation properties of the models, a set of models was computed using \texttt{MESA}
version r10398. To look at the effects in different mass regimes, models were computed for 3.25\msol and 12M$_{\odot}$, corresponding to a typical SPB and $\beta$ Cep star, respectively. We consider regimes for values of $\Dcbm (\rcc < r < \rcbm)$ and envelope mixing $\Dmix(r>\rcbm)$ that are typical for B stars, as found from asteroseismology \citep{2015A&A...580A..27M,2016ApJ...823..130M,2018MNRAS.478.2243S}.
Because of the uncertain physics in the pre-main sequence computations, for example the possible occurrence of intermediate convective zones and how convective boundary mixing processes would influence the pre-main sequence evolution, we opted only to include the core boundary mixing processes from the start of the main sequence, when the fusion of hydrogen in the convective core has started.
Hence, for each stellar mass, one pre-main sequence track was computed for which no overshooting was included. The main sequence evolution for all the different mixing prescriptions was started from this same zero-age main sequence (ZAMS) model.

The \texttt{MESA} models were all calculated using the Ledoux criterion for convection without allowing for semi-convection. The mixing length theory as developed by \citet{1968pss..book.....C} was used, where $\alpha_{\text{mlt}}=2.0$ is the value for the mixing length parameter.
The parameter $f_0$, which determines where the transition from core to near-core mixing is made, is set as $f_0=0.005$ for all cases discussed in \cref{sect:mixing_prescriptions}.
The models were made using the OP opacity tables \citep{2005MNRAS.362L...1S} and
the standard chemical mixture of OB stars in the solar neighbourhood by
\citet{2012A&A...539A.143N,2013EAS....63...13P}, for which an initial hydrogen
content $X_{\rm ini}=0.71$ and an initial metallicity $Z_{\rm ini}=0.014$. The
constant amount of mixing in the radiative envelope is set to
$\Dmix=5$cm$^2$s$^{-1}$.
Such envelope mixing represents the joint effect of macroscopic mixing due to rotation, meridional circulation, internal gravity waves, and magnetism, for example. \citep[e.g.][]{2013LNP...865...23M,doi:10.1146/annurev-astro-091918-104359}, and varies from star to star. Envelope mixing
was found to be a necessary ingredient to model the mode trapping properties of B stars \citep[e.g.][]{2010Natur.464..259D,2015A&A...580A..27M,2016ApJ...823..130M}. We took a value of 5cm$^2$s$^{-1}$ for this paper, in line with the {\it Kepler\/} results of the SPB rotating at 25\% of its
critical rate, as a prototypical case.

Apart from the different near-core mixing prescription and temperature gradient,
the only difference between the \texttt{MESA} models for the different
prescriptions lies in the choice of the free parameter $\fcbm$ and the
inclusion of $\alpha$ in case of the extended convective penetration.
These parameters were chosen in such a way (see \Cref{table:mix_param}) that
the extent of the mixing region $\rcbm$ was approximately equal for each case,
their differences in mass coordinate being smaller than 0.5\% at the considered
$X_{\rm c}$ values during their main sequence evolution. 

\begin{table}[ht]
\caption{Choice of $\fcbm$ and $\alpha$ for the different models. }
\label{table:mix_param} 
\centering                     
\begin{tabular}{l| c c}     
\hline
  &3.25\msol & 12\msol \\  
\hline                      
Diffusive exponential: $\fcbm$ & 0.029  & 0.029 \\    
Diffusive Gumbel: $\fcbm$ & 0.014 &  0.014  \\
Convective Gumbel: $\fcbm$ & 0.016 &  0.015  \\
Extended convective penetration: $\fcbm$ & 0.029 &  0.029   \\
Extended convective penetration: \: $\alpha$ & 0.04 &  0.03   \\
\hline                  
\end{tabular}
    \begin{flushleft}
      \textbf{Notes.} $\alpha=0$ for all models except in the case of extended convective penetration.
    \end{flushleft}
\end{table}

To be able to make a comparison between models with different mixing prescriptions, they are required to be at the same evolutionary stage. Therefore, all models along the evolutionary track are considered at specific values of the central hydrogen content $X_{\rm c}$, while ensuring that their $X_{\rm c}$ value differs from the specified value by less than 0.0005. This entails that the $X_{\rm c}$ difference between models of the various prescriptions is always smaller than 0.001.
The detailed \texttt{MESA} set-up is provided through the link in \cref{appendix:inlist}.

\subsection{\texttt{GYRE}}
The stellar oscillation code \texttt{GYRE}
\citep{2013MNRAS.435.3406T,2018MNRAS.475..879T}, version 5.2, was employed to
compute the pulsation mode properties of the stellar models. In this work, the
adiabatic approximation was used, and a rotation rate of 25\% of the critical
Roche rotation velocity was included for the computation of the pulsation
modes. The g modes are treated in the TAR following
\citet{2018MNRAS.475..879T}, while the p modes are computed from a perturbative
approach \citep[see][for details]{2019A&A...624A..75A}.

Given that the majority of g modes in {\it Kepler\/} data of intermediate-mass stars are prograde dipole modes \citep{2013MNRAS.432..822W,2015ApJS..218...27V,2016ApJ...823..130M,2017MNRAS.465.2294O,2017A&A...598A..74P,2018MNRAS.478.2243S}, we restricted our computations to such modes, i.e. $(l,m)=(1,1)$.  The radial orders $n_{pg}$ vary depending on the mass of the stellar models to cover the range of radial orders typically observed in B stars. For the models with a mass of 3.25M$_{\odot}$, $n_{pg}$ ranges from -50 to -1, whereas for the models with a mass of 12M$_{\odot}$,
$n_{pg}$ ranges from -10 to +5. The \texttt{GYRE} code uses negative and positive values of the radial order to indicate g- and p modes, respectively.  The final \texttt{GYRE} inlist is provided through the link in \cref{appendix:inlist}.

\section{Results for the mode differences}

\begin{figure*}[ht] 
    \centering
        \begin{subfigure}{0.497\hsize}
        \includegraphics[width=\hsize]{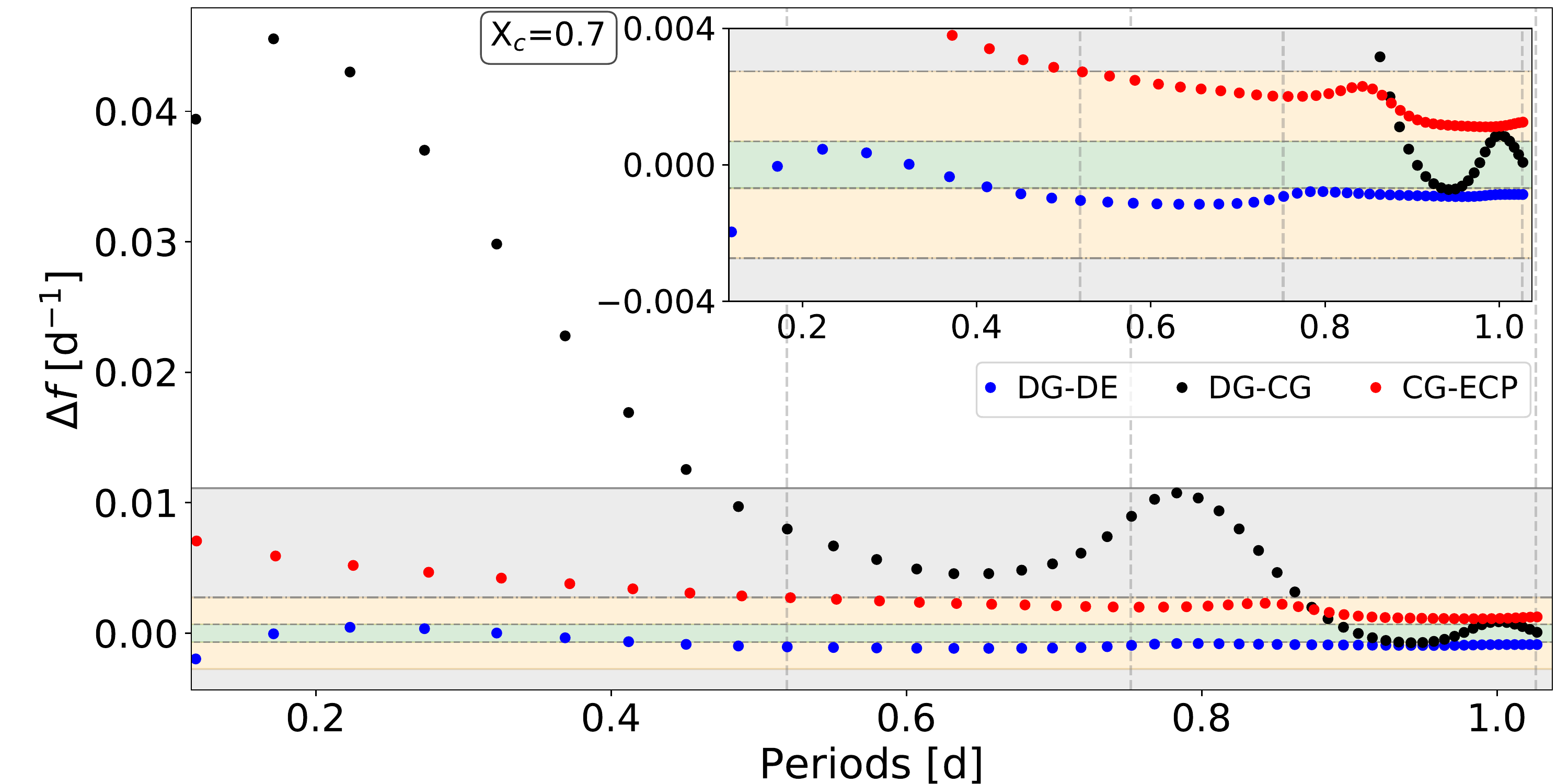}
        \caption{}
        \label{fig:dP_M3Xc7}
        \end{subfigure}
        \begin{subfigure}{0.497\hsize}
        \includegraphics[width=\hsize]{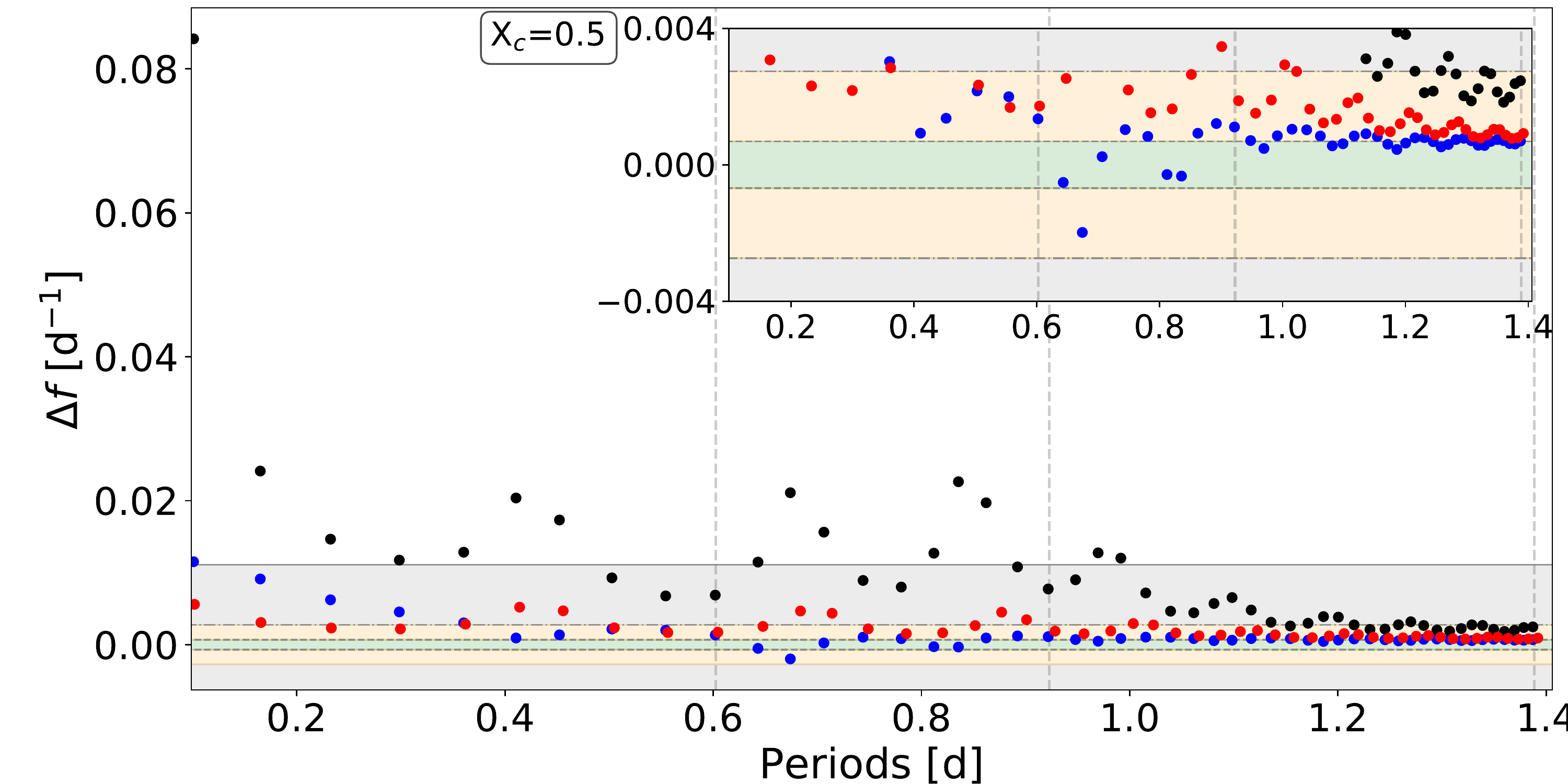}
        \caption{}
        \label{fig:dP_M3Xc5}
        \end{subfigure}     \\
        \begin{subfigure}{0.497\hsize}
        \includegraphics[width=\hsize]{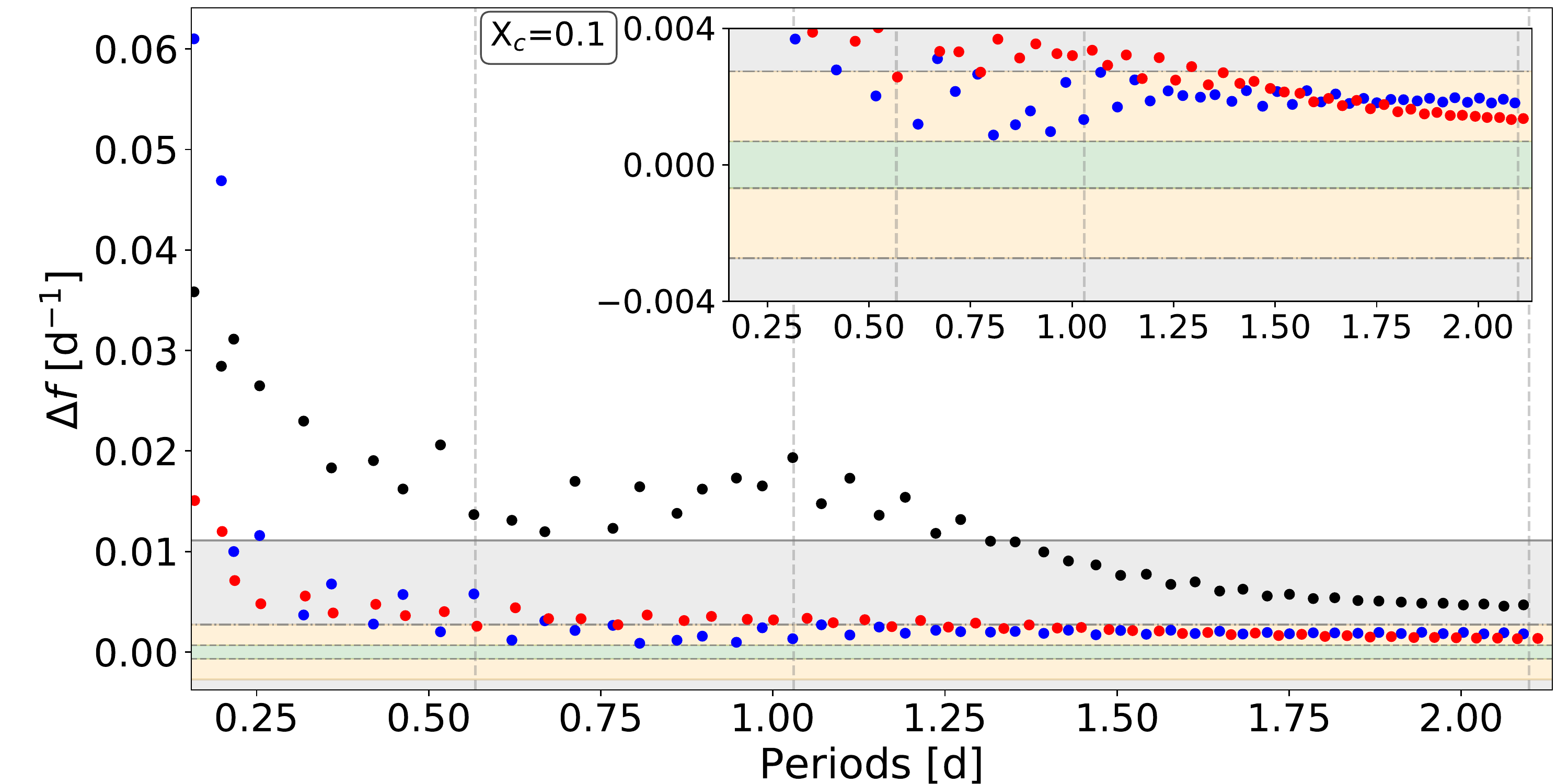}
        \caption{}
        \label{fig:dP_M3Xc1}
        \end{subfigure}
        \begin{subfigure}{0.497\hsize}
        \includegraphics[width=\hsize]{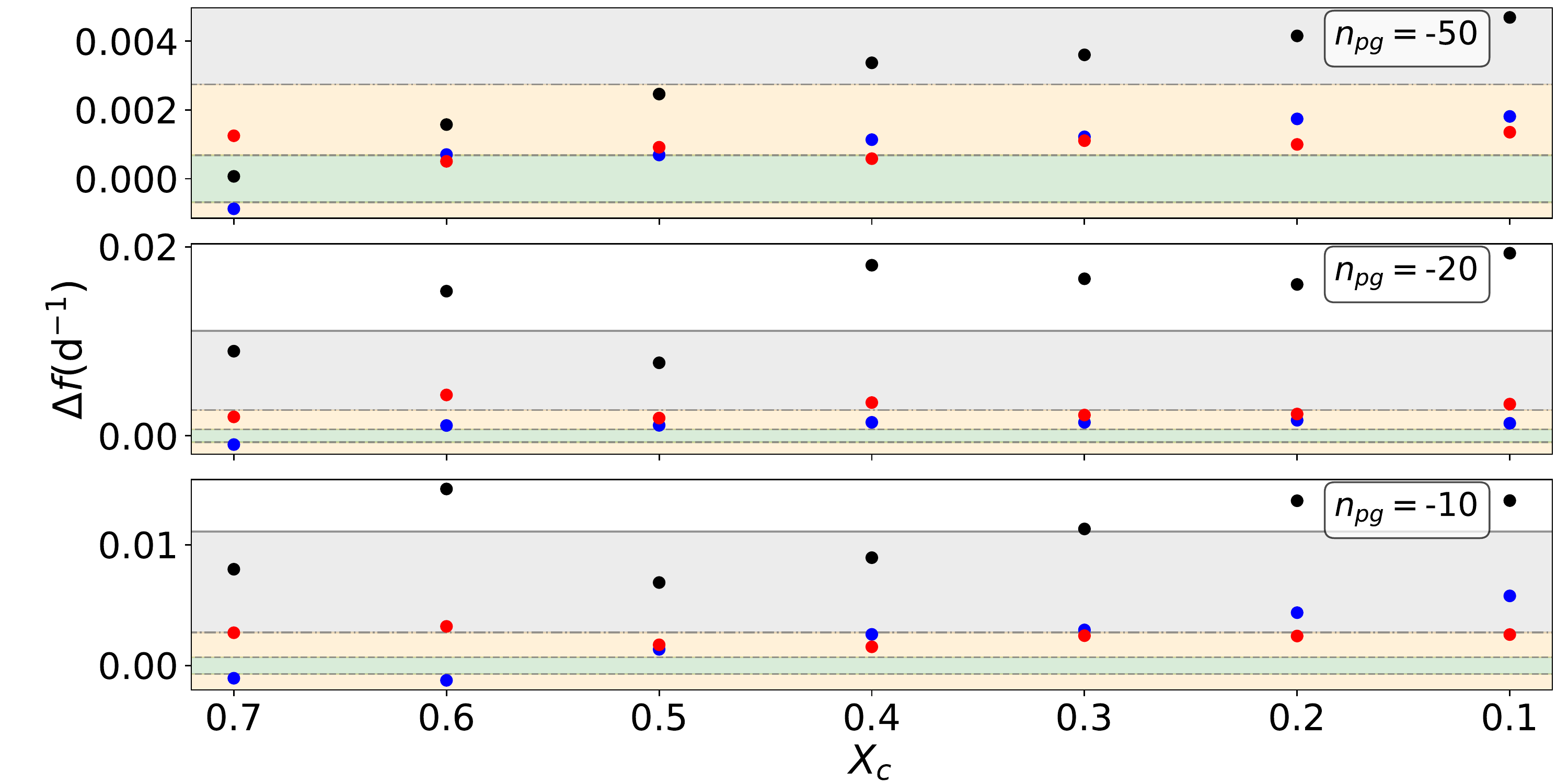}
        \caption{}
        \label{fig:dfEvol_M3}
        \end{subfigure}   
    \caption{\textbf{Panels (a), (b), and (c)}: difference in mode frequencies of radial orders $n_{pg} \in [-50, -1]$ of 3.25\msol models between the given mixing prescriptions: diffusive Gumbel (DG), diffusive exponential (DE), convective Gumbel (CG), and extended convective penetration (ECP). 
    The horizontal lines delineate coloured regions that correspond to the Rayleigh limits of light curves of different lengths: solid line and grey area indicate 90-day-long data (K2), the dash-dotted line and yellow area indicate  one-year-long light curves (TESS) and the dashed line and green area indicate four-year-long light curves (\textit{Kepler}). The vertical dashed lines indicate the modes with $n_{pg}=-10$, -20, and -50 from left to right.
    The inset shows a zoomed version.
    \textbf{Panel (d)}: evolution of the frequency differences of certain modes along the main sequence evolution at various core hydrogen fractions.}
    \label{fig:df_M3}
\end{figure*}

\begin{figure*}[ht]  
    \centering
        \begin{subfigure}{0.497\hsize}
        \includegraphics[width=\hsize]{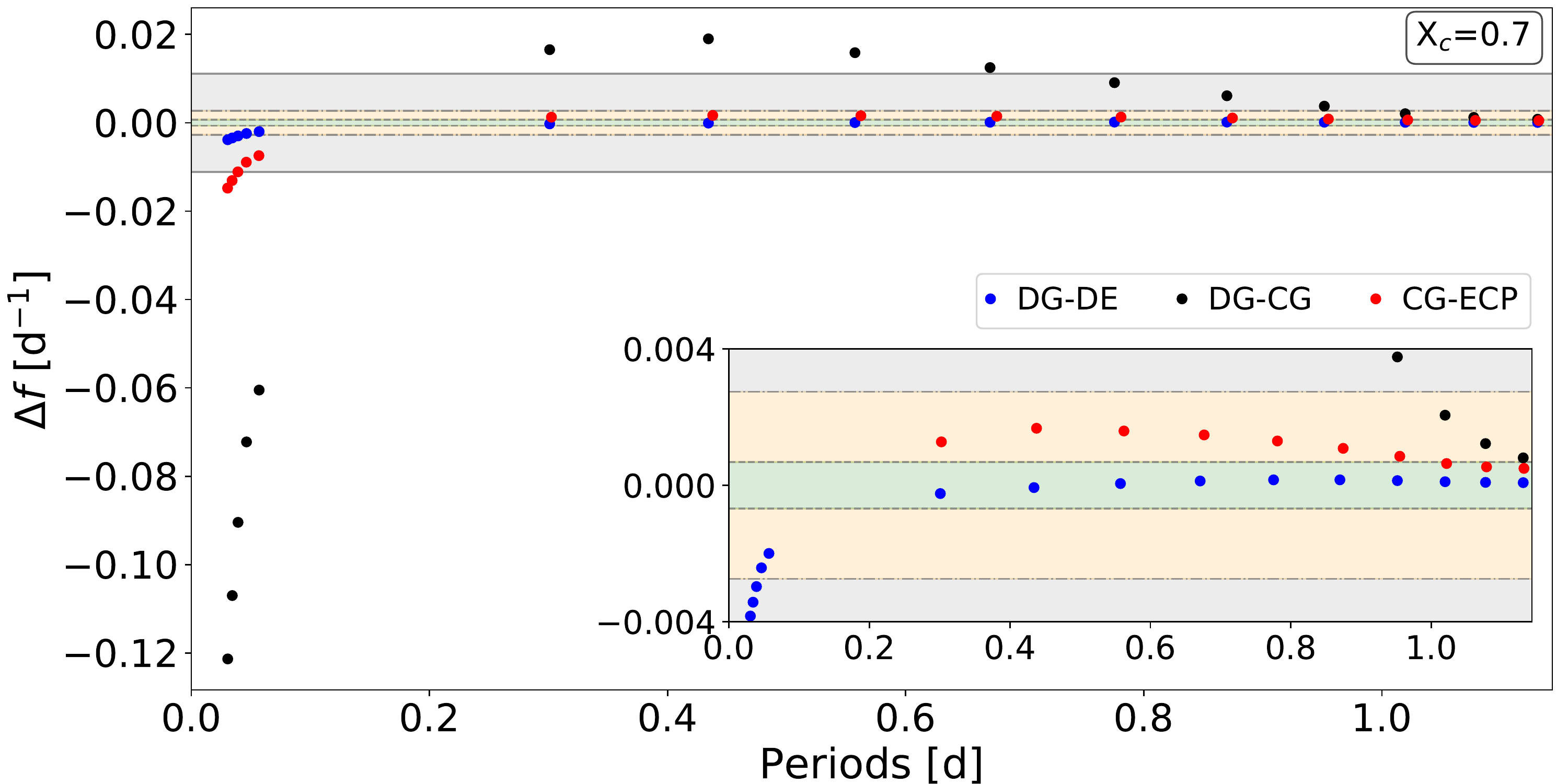}
        \caption{}
        \label{fig:dP_M12Xc7}
        \end{subfigure}
        \begin{subfigure}{0.497\hsize}
        \includegraphics[width=\hsize]{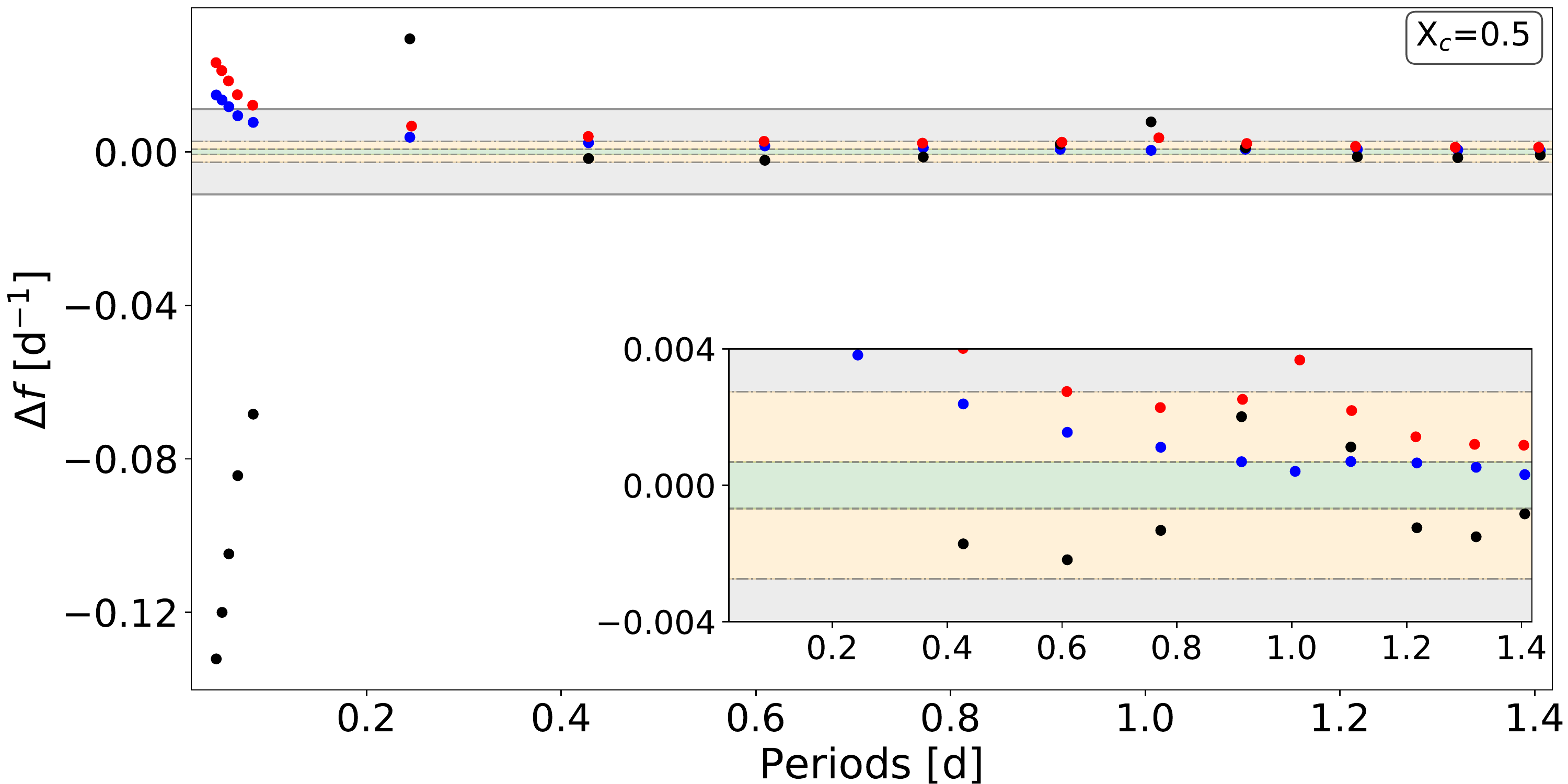}
        \caption{}
        \label{fig:dP_M12Xc5}
        \end{subfigure}     \\
        \begin{subfigure}{0.497\hsize}
        \includegraphics[width=\hsize]{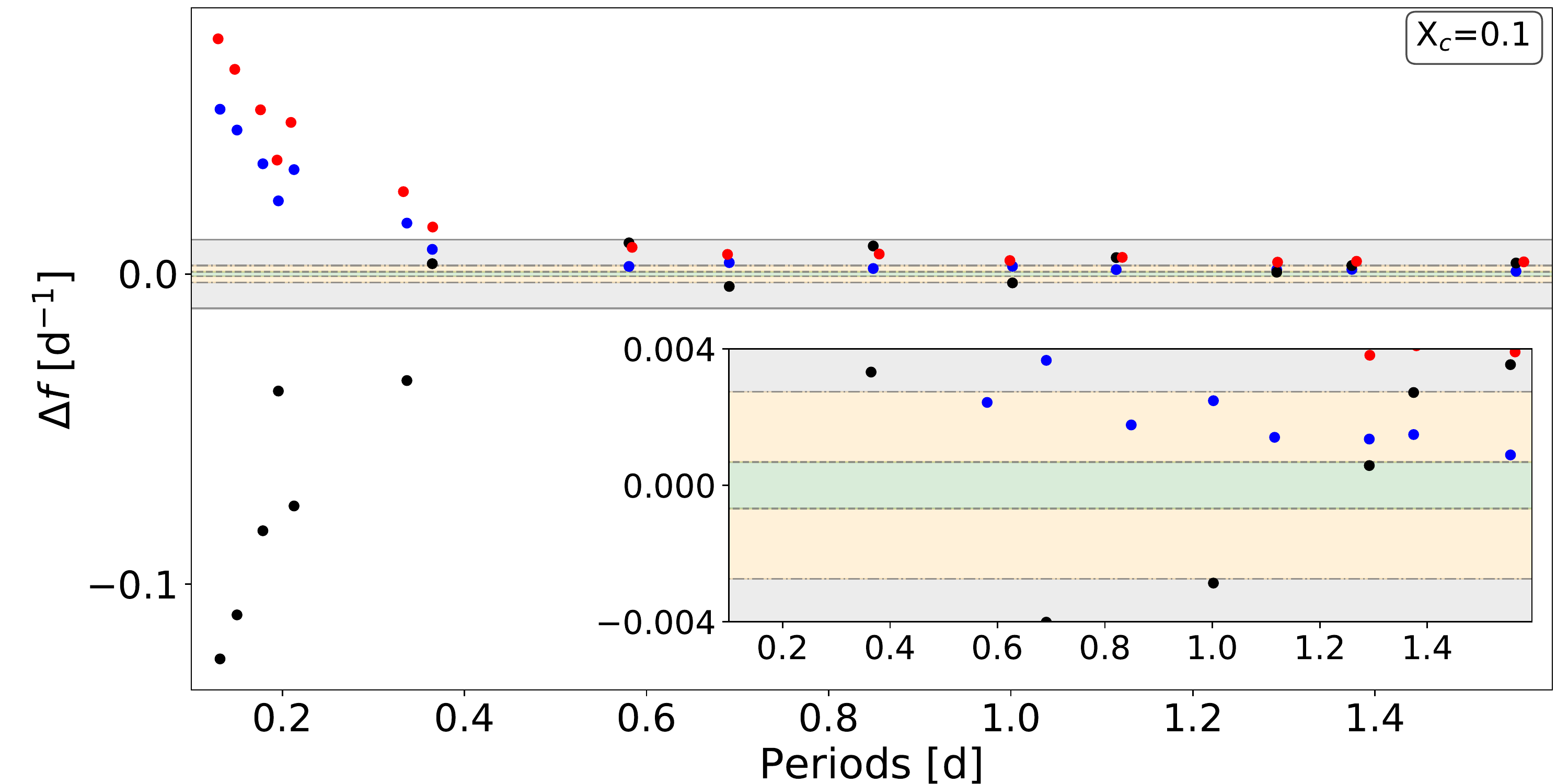}
        \caption{}
        \label{fig:dP_M12Xc1}
        \end{subfigure}
        \begin{subfigure}{0.497\hsize}
        \includegraphics[width=\hsize]{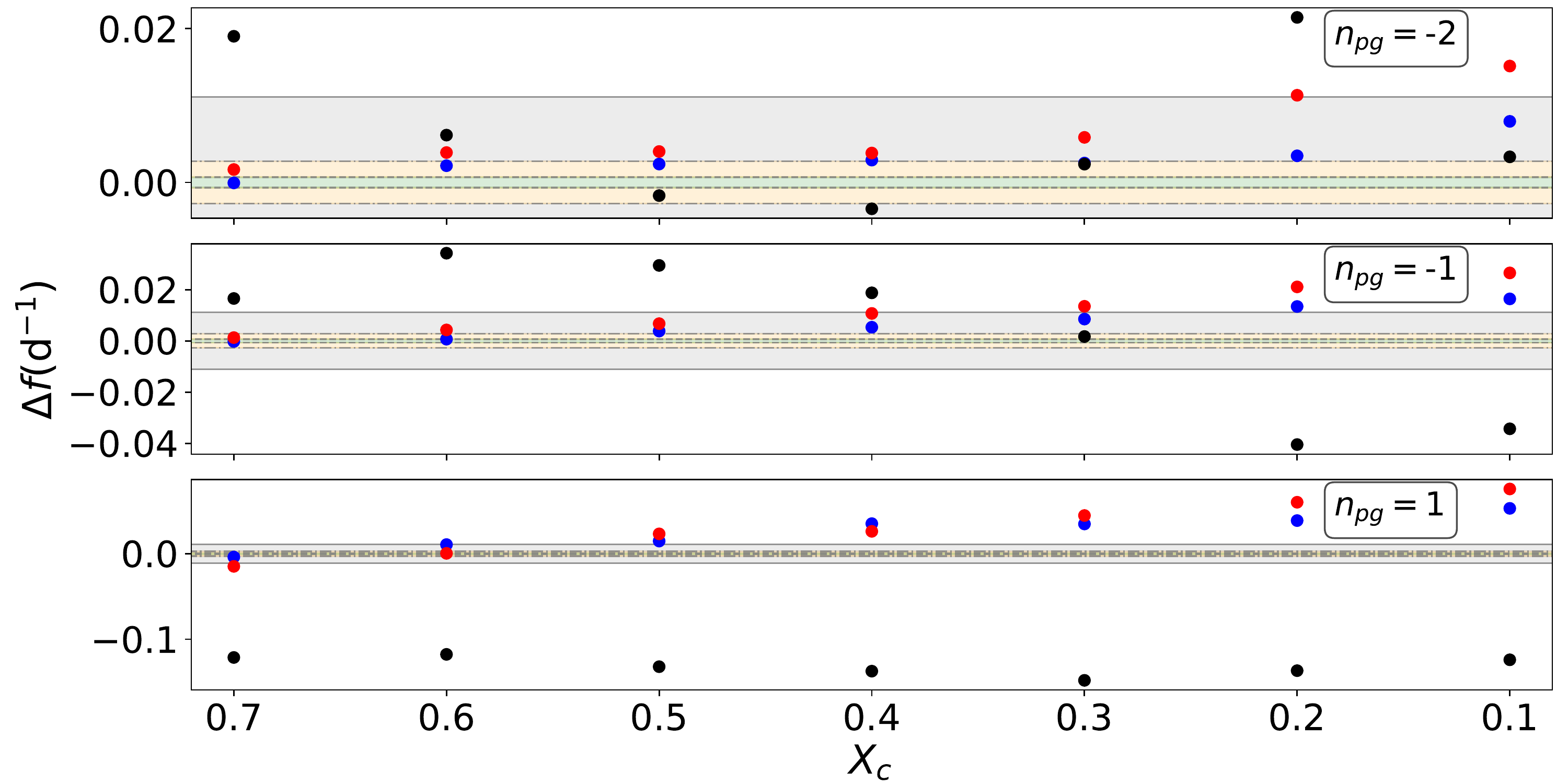}
        \caption{}
        \label{fig:dfEvol_M12}
        \end{subfigure}   
    \caption{Same as \cref{fig:df_M3}, but for radial orders $n_{pg} \in [-10, 5]$ of 12\msol models.}
    \label{fig:df_M12}
\end{figure*}

To illustrate the effects of the different mixing prescriptions, \cref{fig:dP_M3Xc7,fig:dP_M3Xc5,fig:dP_M3Xc1,fig:dP_M12Xc7,fig:dP_M12Xc5,fig:dP_M12Xc1} show the frequency differences $\Delta f = f_n-f'_n$ between models with the different mixing prescriptions for the masses 3.25\msol and 12\msol, respectively. In this case, $f_n$ and $f'_n$ are the frequencies of modes with radial order $n_{pg}$, where $|n_{pg}|$ increases with increasing period for g modes and with decreasing period for p modes. The p modes in \cref{fig:dP_M12Xc7,fig:dP_M12Xc5,fig:dP_M12Xc1} are the five points with the lowest periods. \Cref{fig:dfEvol_M3,fig:dfEvol_M12} illustrate the evolution of these frequency differences for certain modes at various values of $X_{\rm c}$ along the main sequence evolution. A comparison is made between the frequencies obtained from the diffusive and convective Gumbel profiles to investigate the influence of the change in temperature gradient in the overshoot zone. 
Additionally, comparisons are made between the two diffusive models, and between the two convective models, to study the effect of the change in the functional form of the mixing coefficient. 

Theoretically computed oscillation-mode frequencies may give exact numbers, but observational data suffer from the limited resolving power of the data, based upon the length of the time base of the observations.
In addition, measured frequency errors are defined by the number of data points and noise properties of the instrument. Since the latter two differ for various surveys, we work with the Rayleigh limit, which is the inverse of the total time base of the data, to assess for at what levels of resolving power we can potentially probe the thermal structure and mixing profiles near the convective core.
Different space missions have had different covered time bases during which they collected the photometric data used in asteroseismology. \textit{Kepler} \citep{2010Sci...327..977B} collected light curves with a time base of four years, and later K2 continued to observe, yielding data sets of 90 days long. Currently, the Transiting Exoplanet Survey Satellite (TESS) \citep{2015JATIS...1a4003R} is gathering data, and the light curves from stars in its continuous viewing zone will span about one year. The Bright Target Explorer  \citep{2014PASP..126..573W} delivers various time bases \citep[e.g.][]{2017A&A...603A..13K,2019MNRAS.485.3544W}.

To determine if a distinction could be made between the theoretically computed $f_n$ and $f'_n$ if they were compared to frequencies extracted from an observed light curve, a conservative approach was taken by comparing $\Delta f$ to the Rayleigh limit.
Three different Rayleigh limits are considered and are shown in \cref{fig:df_M3,fig:df_M12}, corresponding to observed data sets with a length of 90 days, one year and four years, to match the aforementioned space missions. The percentage of the modes that would be observationally distinguishable based upon these Rayleigh limits can be found in \cref{table:percentDGDE,table:percentDGCG,table:percentCGECP}.

\begin{figure}[ht]
  \centering
        \includegraphics[width=\hsize]{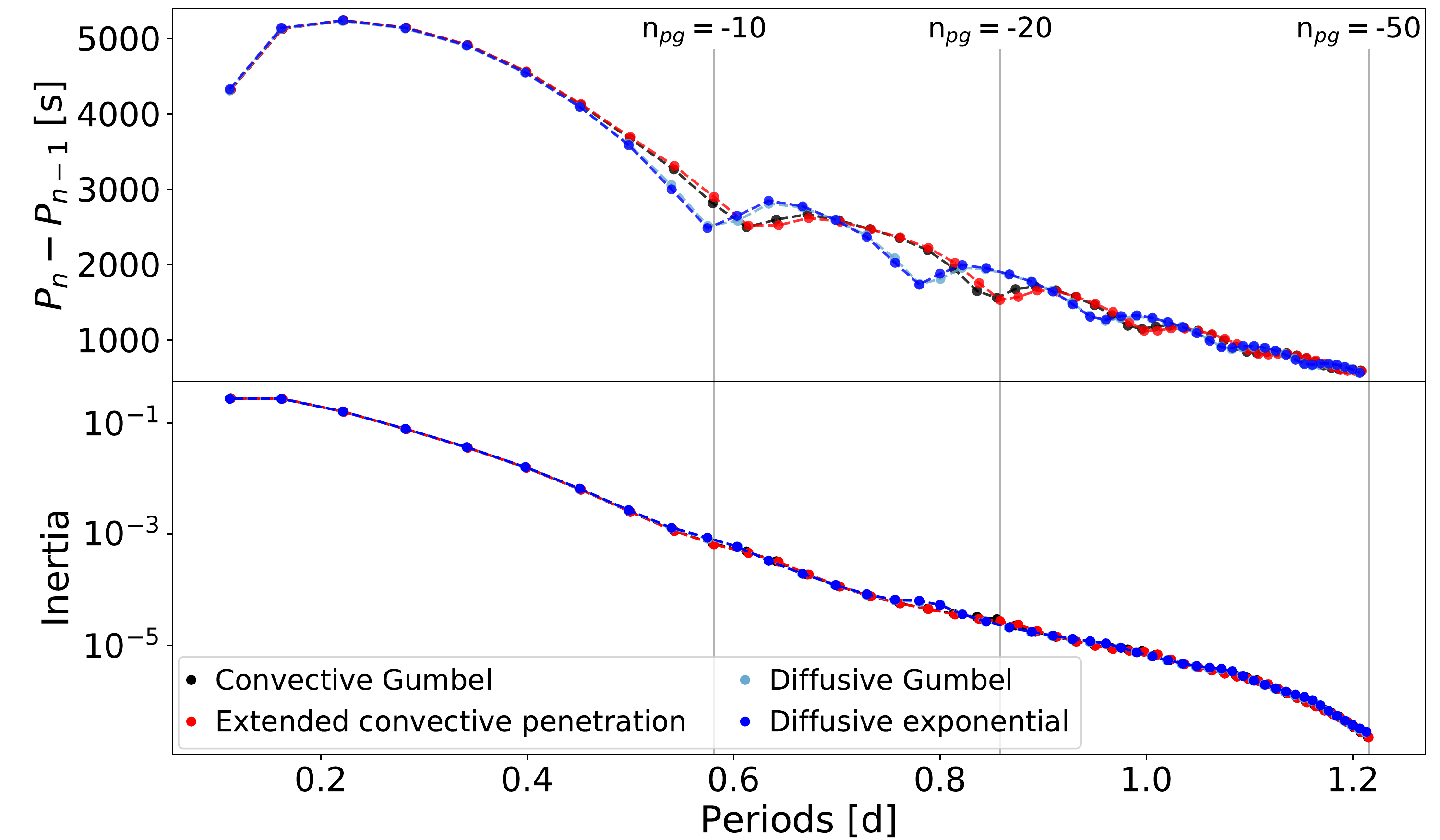}
      \caption{Period spacing and mode inertia of the 3.25\msol models at $X_{\rm c}=0.6$. The vertical grey lines indicate the modes with the given $n_{pg}$.}
         \label{fig:dP_inertia_M3Xc6}
  \end{figure}

The observational diagnostics being used are dominantly the frequencies of the modes. However, once their identification in terms of $l, m,$ and $n_{pg}$ is achieved, we can improve the interpretation by analysing the mode properties, such as their kinetic energy, and the trapping they undergo. We provide the mode inertia, which are proportional to the mode energy, in \cref{fig:dP_inertia_M3Xc6}

\tabcolsep=5pt
\begin{table}[ht]
\caption{Percentage of modes for which the frequency difference between the diffusive Gumbel and diffusive exponential models is larger than the Rayleigh limit for a given data set.} 
\label{table:percentDGDE}      
\centering
\begin{tabular}{c |r r r| r r r| r r r}       
\hline 
$X_{\rm c}$ & \multicolumn{3}{c|} {SPB} & \multicolumn{3}{c|} {$\beta$ Cep g modes} & \multicolumn{3}{c} {$\beta$ Cep p modes}  \\
  & 4yrs & 1yr  & 90d & 4yrs & 1yr  & 90d & 4yrs & 1yr  & 90d \\
\hline 
0.7 & 88  & 0  & 0 & 0   & 0  & 0  & 100 & 60  & 0 \\
0.6 & 76  & 12 & 0 & 50  & 0  & 0  & 100 & 100 & 0 \\
0.5 & 64  & 10 & 2 & 60  & 10 & 0  & 100 & 100 & 60 \\
0.4 & 92  & 12 & 2 & 80  & 20 & 0  & 100 & 100 & 100 \\
0.3 & 98  & 18 & 2 & 90  & 10 & 0  & 100 & 100 & 100 \\
0.2 & 96  & 22 & 4 & 100 & 30 & 10 & 100 & 100 & 100 \\
0.1 & 100 & 20 & 6 & 100 & 30 & 10 & 100 & 100 & 100 \\
\hline
\end{tabular}
\begin{flushleft}
\textbf{Notes.} The first, second, and third number per $X_{\rm c}$ value in each column correspond to the three different Rayleigh limits of 4 years (\textit{Kepler}), 1 year (TESS), and 90 days (K2), respectively. 
\end{flushleft}
\end{table}

\begin{table}[ht]
\caption{Same as \cref{table:percentDGDE}, but for the differences between convective Gumbel and extended convective penetration prescriptions.}             
\label{table:percentCGECP}   
\centering                        
\begin{tabular}{c |r r r| r r r| r r r}     
\hline 
$X_{\rm c}$ & \multicolumn{3}{c|} {SPB} & \multicolumn{3}{c|} {$\beta$ Cep g modes} & \multicolumn{3}{c} {$\beta$ Cep p modes}  \\  
 & 4yrs & 1yr  & 90d & 4yrs & 1yr  & 90d & 4yrs & 1yr  & 90d \\
\hline 
0.7 & 100 & 18 & 0 & 70  & 0   & 0  & 100 & 100 & 40 \\
0.6 & 80  & 18 & 0 & 100 & 20  & 0  & 0   & 0   & 0 \\
0.5 & 100 & 20 & 0 & 100 & 40  & 0  & 100 & 100 & 100 \\
0.4 & 88  & 26 & 0 & 100 & 50  & 0  & 100 & 100 & 100 \\
0.3 & 100 & 26 & 2 & 100 & 80  & 10 & 100 & 100 & 100 \\
0.2 & 100 & 32 & 2 & 100 & 90  & 20 & 100 & 100 & 100 \\
0.1 & 100 & 44 & 4 & 100 & 100 & 20 & 100 & 100 & 100 \\
\hline
\end{tabular}
\end{table}

\begin{table}[ht]
\caption{Same as \cref{table:percentDGDE}, but for the differences between the diffusive Gumbel and convective Gumbel prescriptions.} 
\label{table:percentDGCG} 
\centering             
\begin{tabular}{c |r r r| r r r| r r r} 
\hline 
$X_{\rm c}$ & \multicolumn{3}{c|} {SPB} & \multicolumn{3}{c|} {$\beta$ Cep g modes} & \multicolumn{3}{c} {$\beta$ Cep p modes} \\ 
  & 4yrs & 1yr  & 90d & 4yrs & 1yr  & 90d & 4yrs & 1yr  & 90d \\
\hline 
0.7 & 72  & 56  & 16 & 100 & 70 & 40 & 100 & 100 & 100 \\
0.6 & 100 & 66  & 22 & 80  & 30 & 10 & 100 & 100 & 100 \\
0.5 & 100 & 74  & 30 & 100 & 20 & 10 & 100 & 100 & 100 \\
0.4 & 100 & 100 & 38 & 90  & 40 & 20 & 100 & 100 & 100 \\
0.3 & 100 & 100 & 42 & 100 & 50 & 10 & 100 & 100 & 100 \\
0.2 & 100 & 100 & 48 & 90  & 60 & 20 & 100 & 100 & 100 \\
0.1 & 100 & 100 & 52 & 90  & 80 & 10 & 100 & 100 & 100 \\
\hline
\end{tabular}
\end{table}

\begin{figure*}[ht]  
    \centering
        \begin{subfigure}{0.497\hsize}
        \includegraphics[width=\hsize]{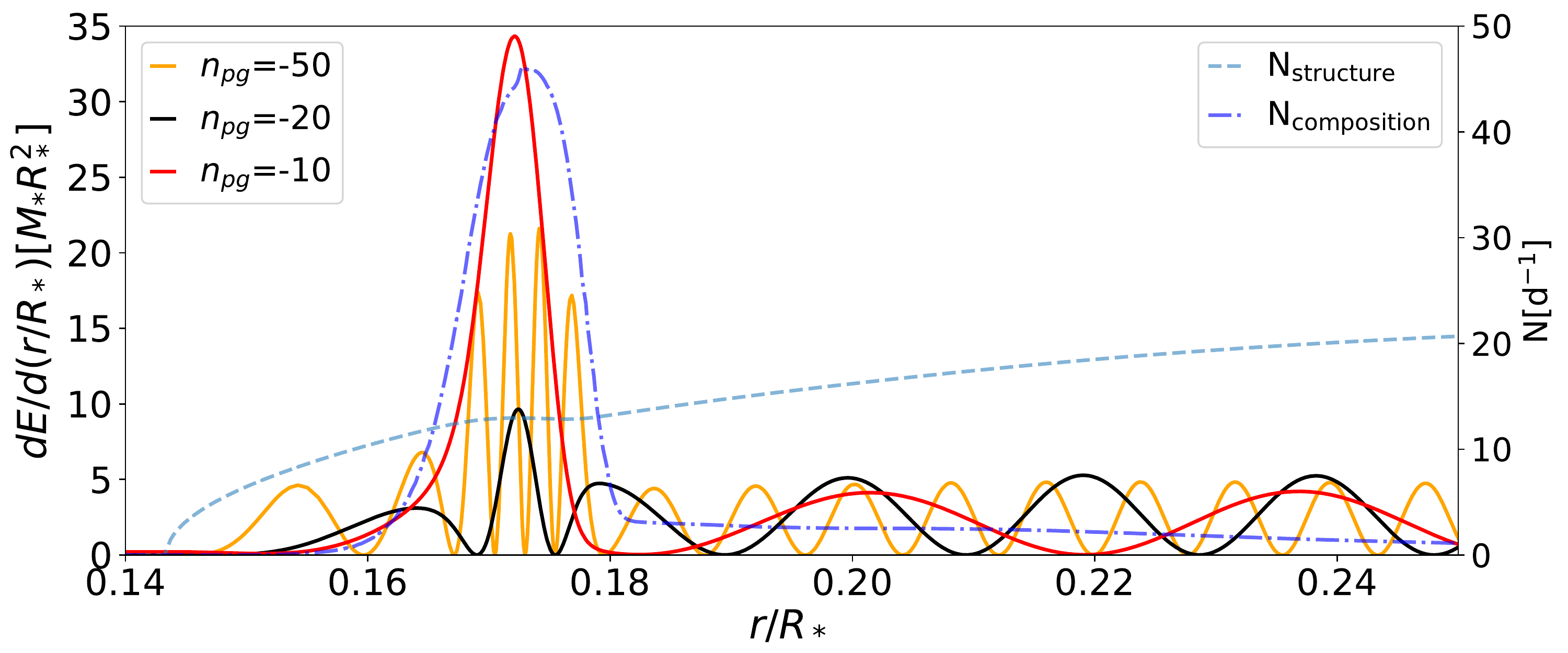}
        \caption{Diffusive exponential}
        \label{fig:inertia_M3Xc6DE}
        \end{subfigure}
        \begin{subfigure}{0.497\hsize}
        \includegraphics[width=\hsize]{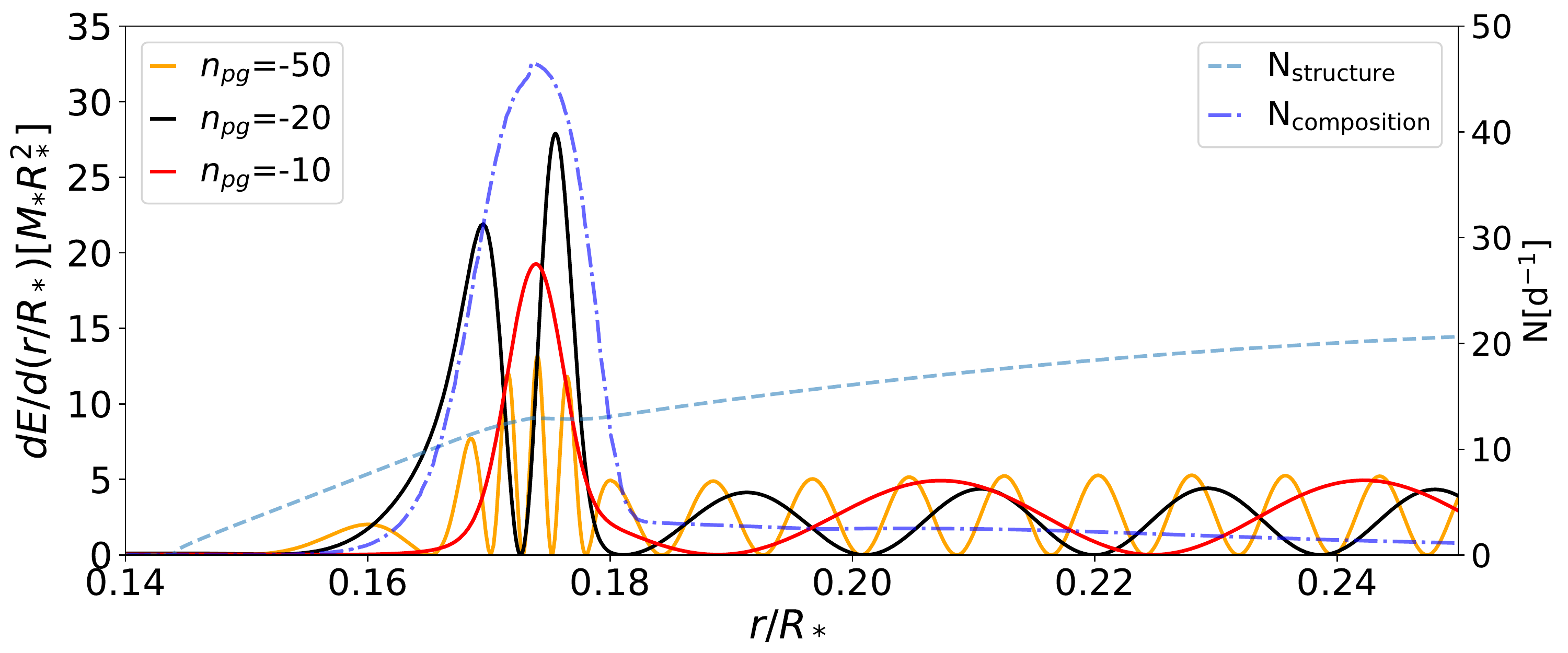}
        \caption{Extended convective penetration}
        \label{fig:inertia_M3Xc6ECP}
        \end{subfigure}   \\        
        \begin{subfigure}{0.497\hsize}
        \includegraphics[width=\hsize]{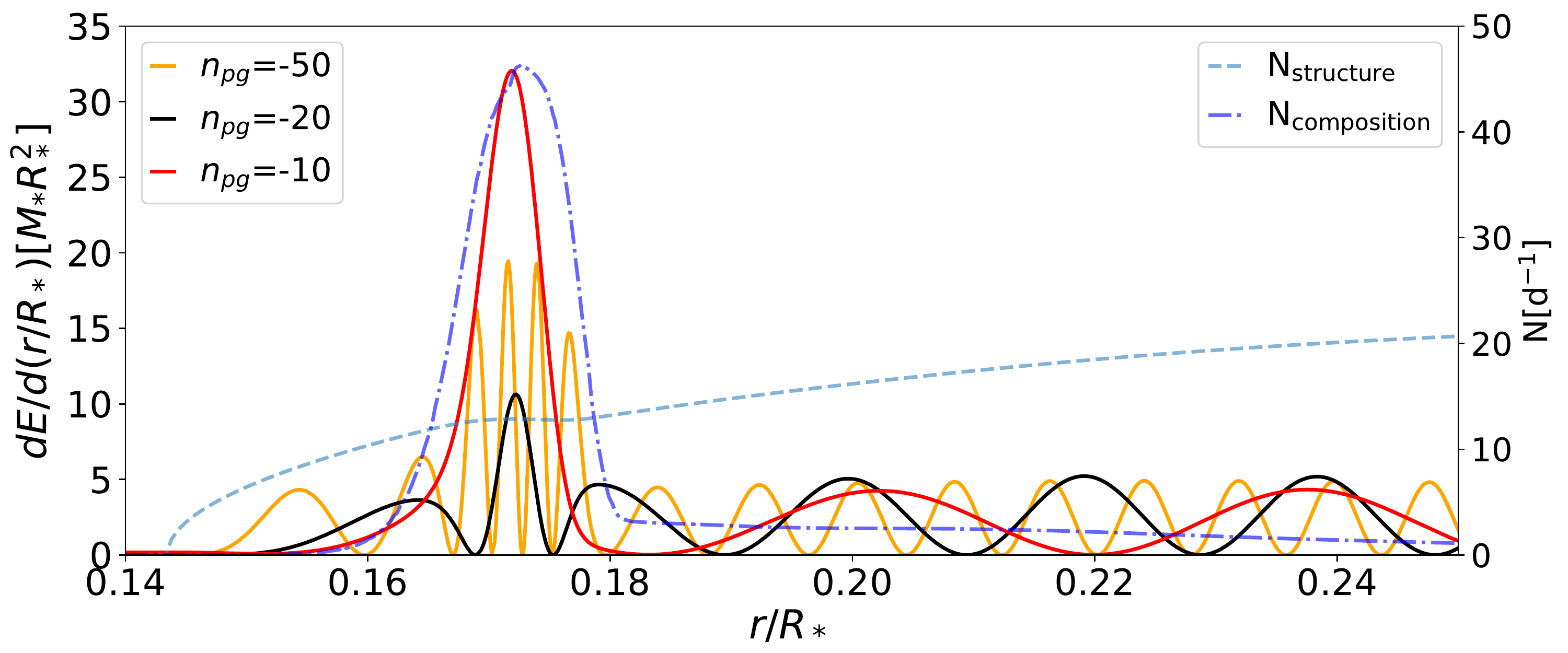}
        \caption{Diffusive Gumbel}
        \label{fig:inertia_M3Xc6DG}
        \end{subfigure} 
        \begin{subfigure}{0.497\hsize}
        \includegraphics[width=\hsize]{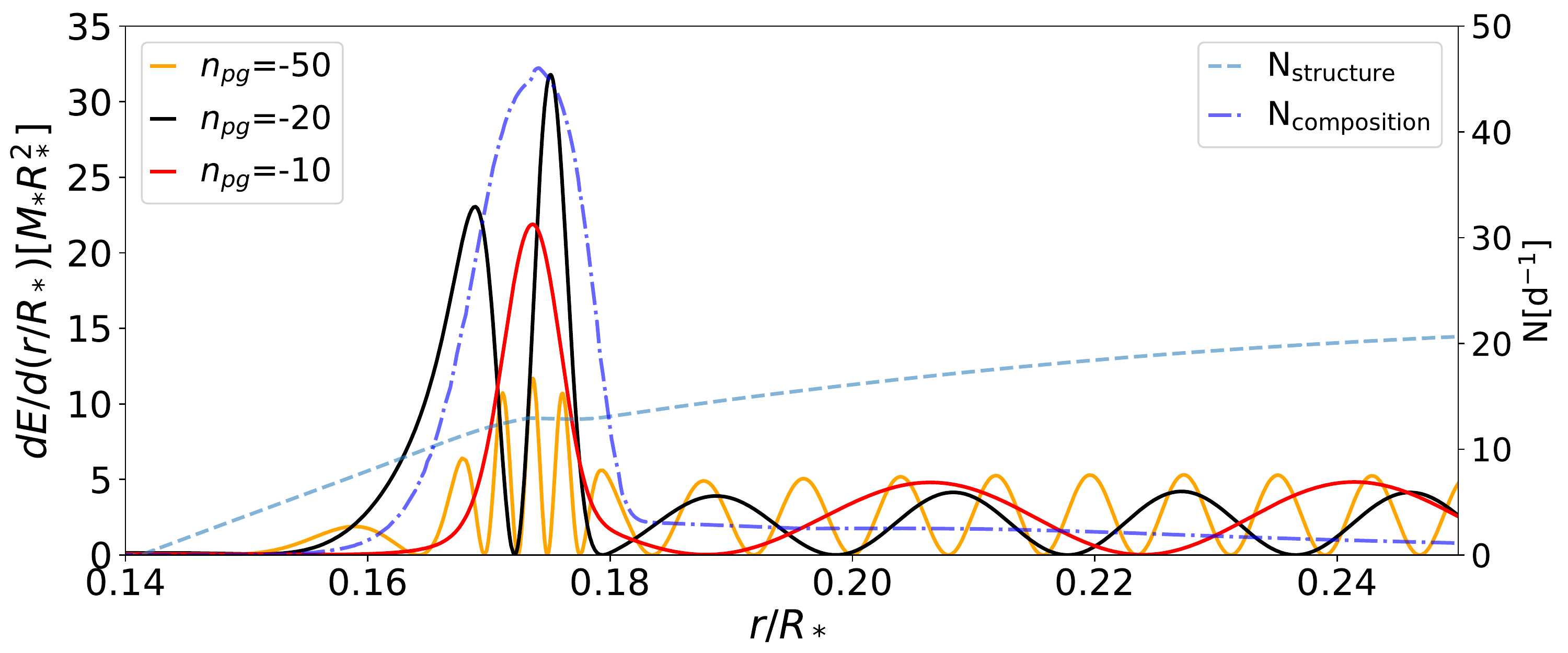}
        \caption{Convective Gumbel}
        \label{fig:inertia_M3Xc6CG}
        \end{subfigure}
    \caption{Differential mode inertia, with E the energy and r/R$_*$ the fractional radius, alongside the components of the Brunt-V\"ais\"al\"a frequency of 3.25\msol models at $X_{\rm c}=0.6$.}
    \label{fig:inertia_M3Xc6}
\end{figure*}

\begin{figure*}[ht]  
    \centering
        \begin{subfigure}{0.497\hsize}
        \includegraphics[width=\hsize]{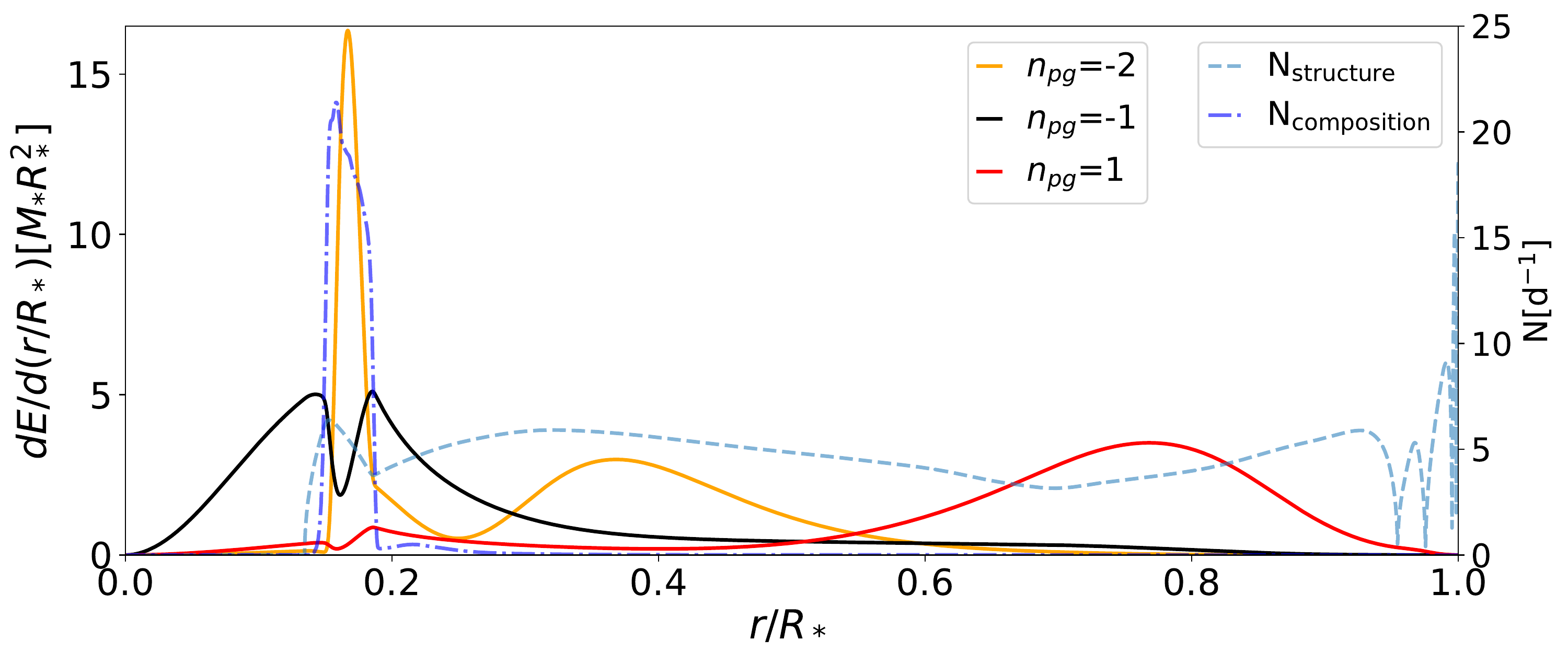}
        \caption{Diffusive exponential}
        \label{fig:inertia_M12Xc3DE}
        \end{subfigure}
        \begin{subfigure}{0.497\hsize}
        \includegraphics[width=\hsize]{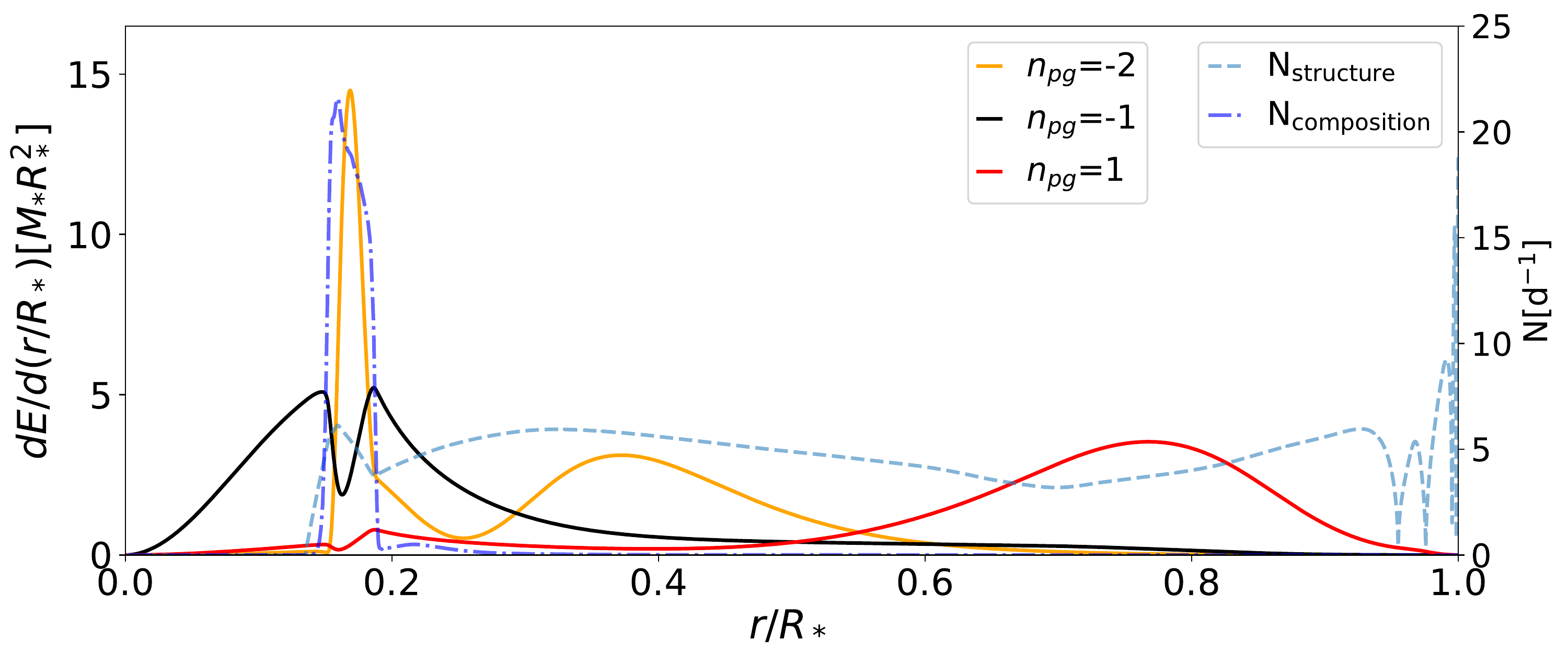}
        \caption{Extended convective penetration}
        \label{fig:inertia_M12Xc3ECP}
        \end{subfigure}   \\
        \begin{subfigure}{0.497\hsize}
        \includegraphics[width=\hsize]{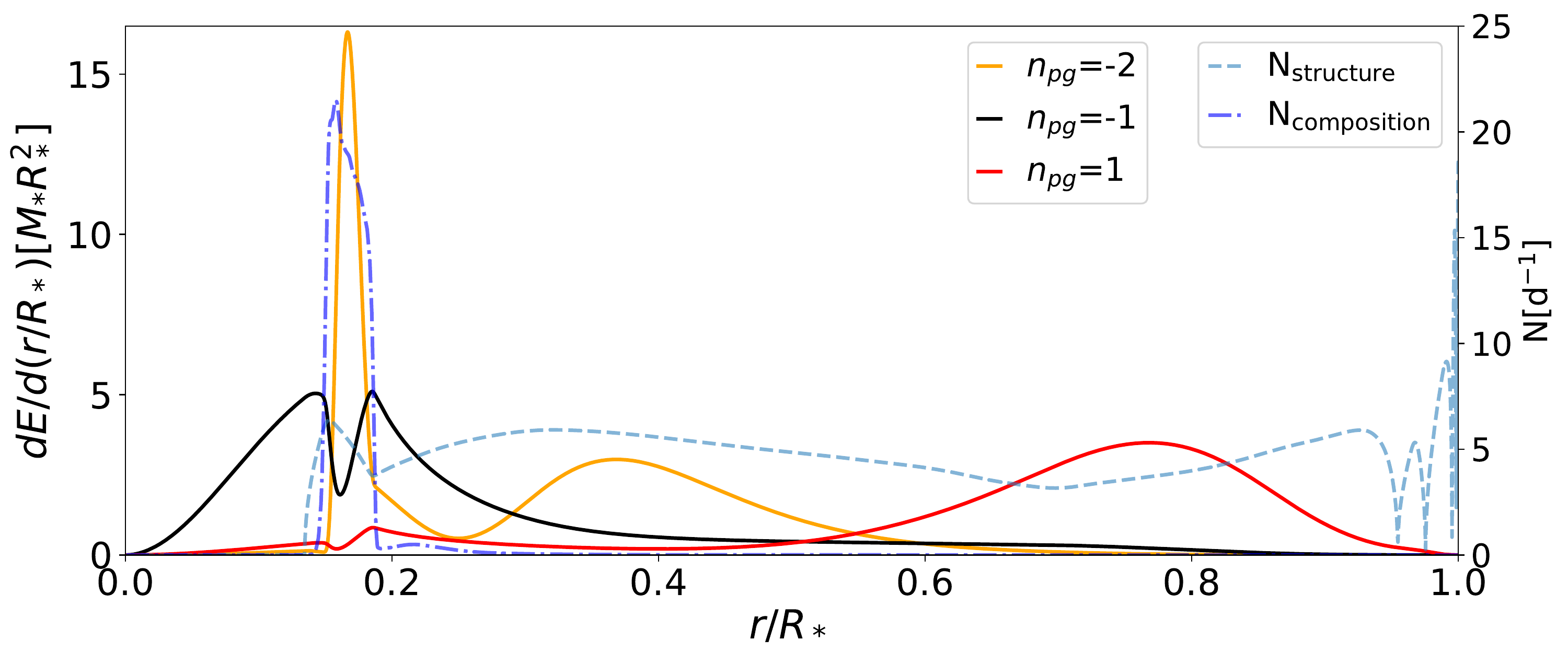}
        \caption{Diffusive Gumbel}
        \label{fig:inertia_M12Xc3DG}
        \end{subfigure}
        \begin{subfigure}{0.497\hsize}
        \includegraphics[width=\hsize]{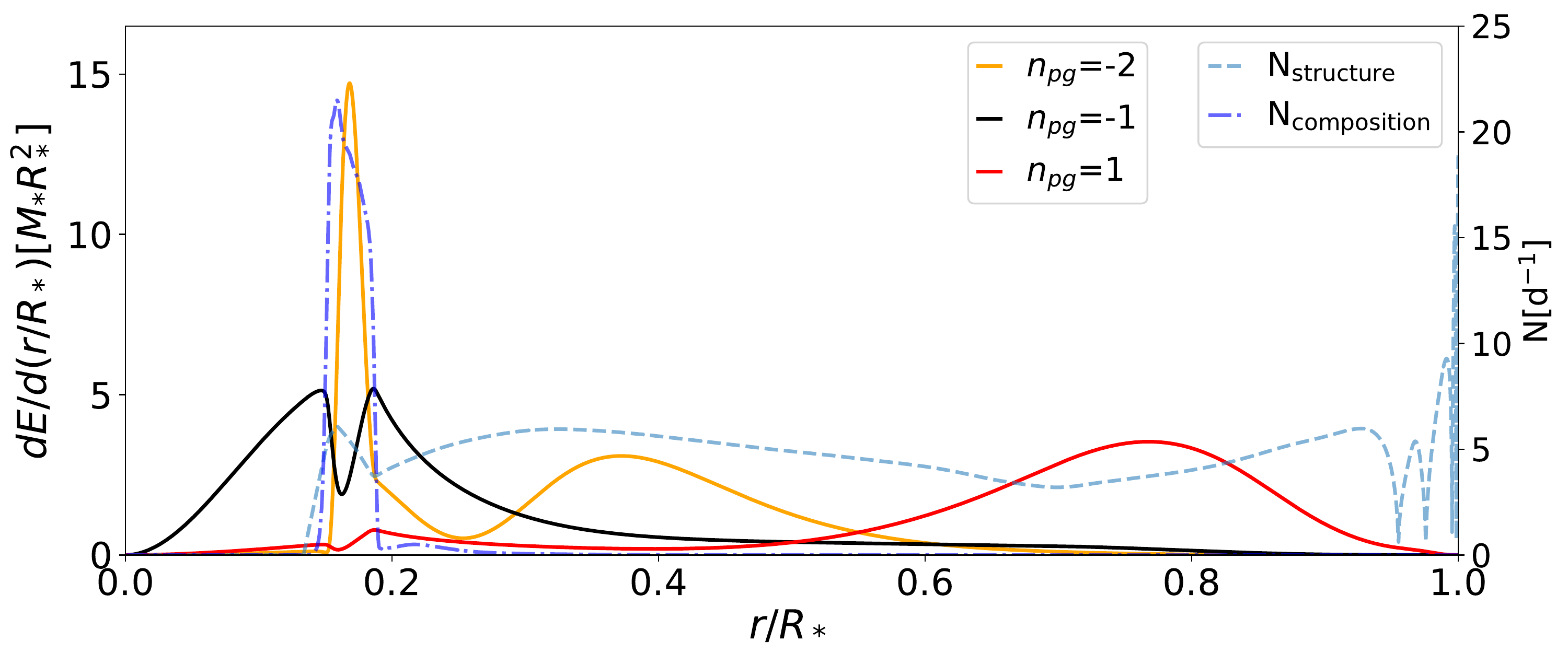}
        \caption{Convective Gumbel}
        \label{fig:inertia_M12Xc3CG}
        \end{subfigure}
    \caption{Same as \cref{fig:inertia_M3Xc6}, but for 12\msol models at $X_{\rm c}=0.3$.}
    \label{fig:inertia_M12Xc3}
\end{figure*}

\subsection{SPB stars}

The 3.25\msol models correspond to SPB stars.
When using a 90-day-long data set, we can only discern between the diffusive and convective profiles based on the frequency values using low order g modes.  The evolutionary stage of the models determines which mode orders are considered low enough.

The mode inertia in \cref{fig:dP_inertia_M3Xc6} are comparable between the different models. The two diffusive models are very similar to each other both in mode inertia and period spacing pattern, just as the two convective models. However, the radial orders for which mode trapping occurs differ between the diffusive and convective models. \Cref{fig:dP_inertia_M3Xc6} clearly shows that trapping occurs for, for example $n_{pg}=-20,$ in the convective models through the characteristic dip in the period spacing and a slightly higher mode inertia. The trapping at $n_{pg}=-20$ is absent in the diffusive models, but occurs at a different radial order.
The occurrence of mode trapping for specific radial orders hence is a way to probe the temperature structure in the near-core region. Models near the ZAMS which have yet to develop the chemical gradient responsible for the mode trapping are therefore more difficult to discern.

\begin{figure}[ht]
  \centering
        \includegraphics[width=\hsize]{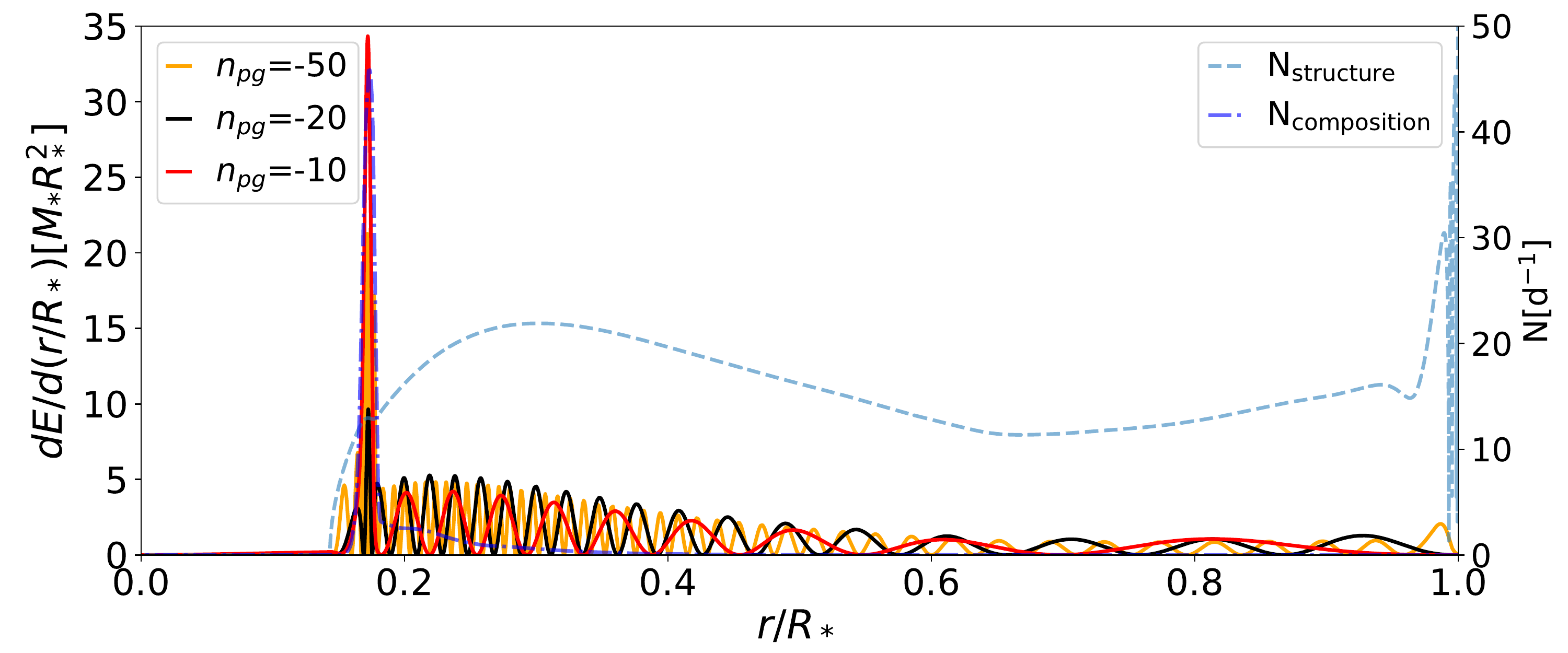}
      \caption{Differential mode inertia and components of the Brunt-V\"ais\"al\"a frequency of a 3.25\msol model at $X_{\rm c}=0.6$.}
         \label{fig:inertia_M3Xc6DE_fullRadius}
  \end{figure}

The differential mode inertia in \cref{fig:inertia_M3Xc6} shows that the differences in probing capacity are largest between convective (\cref{fig:inertia_M3Xc6CG,fig:inertia_M3Xc6ECP}) and diffusive (\cref{fig:inertia_M3Xc6DE,fig:inertia_M3Xc6DG}) models, and that they are larger for the lower and intermediate order modes ($n_{pg}$ of -10 and -20) than for the high order modes ($n_{pg}$ of -50).
The modes are most sensitive to the near-core region where the composition term of the Brunt-V\"ais\"al\"a frequency (N$_{\text{composition}}=\frac{g^2 \rho}{P} \nabla_{\mu}$) dominates over the thermal structure term (N$_{\text{structure}}=\frac{g^2 \rho}{P} (\nad-\n$)).
As can be seen from \cref{fig:inertia_M3Xc6DE_fullRadius}, where the differential inertia across the full radius of the star is shown, the g modes are not very sensitive to the stellar envelope. 

In general, it becomes more difficult to make the distinction between the frequency values for higher order g modes because of the higher mode density at longer periods.
For the most evolved stages and the lowest order g modes, a few of the frequency differences between the models with the same temperature structures become larger than the Rayleigh limit of a 90-day data set. However, it would be very challenging to discern these profiles observationally, since this is only the case for very few modes.

From data with a one-year-long time base, we can distinguish between diffusive and convective profiles for all but some very high order g modes in the early stages of the evolution. For some low order modes in certain evolutionary stages, the frequency differences caused by the changes in the mixing profiles become significant as well.

For a set of data that is four years long, there is a frequency difference larger than the Rayleigh limit for almost all modes. Only some of the differences caused by the change in mixing coefficient remain smaller than the Rayleigh limit in the less evolved models, as can be seen in \cref{table:percentDGDE,table:percentDGCG,table:percentCGECP}.
Both \cref{fig:dfEvol_M3} and the percentages in these tables indicate that the effects on the frequencies become larger for more evolved models, making the distinction easier as the models get more evolved.
\newline

\subsection{\texorpdfstring{$\beta$}{TEXT} Cep stars}

Considering the 12\msol models, representing $\beta$ Cep stars, both
low-order p and g modes are examined because it has become clear from space
photometry that such pulsators exhibit both these types of modes simultaneously
\citep[e.g.][]{2017MNRAS.464.2249H,2019MNRAS.485.3544W}.  Typically, modes with
$l=0, 1, 2$ and $n_{pg}=-2, -1, +1$ are observed for such stars
\citep[e.g.][]{2003Sci...300.1926A}.  As can be seen from the differential mode
inertia in \cref{fig:inertia_M12Xc3}, the p modes are most sensitive to the
stellar envelope in contrast to the g modes.  The frequency differences of the
p modes can therefore be linked to the stellar radii listed in
\cref{table:radii_Bcep}. At the ZAMS, the radii of models with the diffusive
exponential and Gumbel profiles are slightly larger than models based on the
convective Gumbel and extended convective penetration profiles. The frequencies
of the latter two are therefore higher than those of the former two, as can
be seen by the negative $\Delta f$ in \cref{fig:dfEvol_M12}.

The frequency differences become smallest around $X_{\rm c}=0.6$ when comparing
the convective models, and at an $X_{\rm c}$ slightly lower than 0.7 for the
diffusive models. These $X_{\rm c}$ values correspond to the stages at which the
radii are approximately equal between the compared models, explaining why there are only minor differences, smaller than the four-year Rayleigh limit, between the p-mode frequencies at that evolutionary phase (cf.\ Table\,\ref{table:percentCGECP}). At more evolved stages, the radii of the diffusive exponential and extended convective penetrative models are larger than those of the diffusive and convective Gumbel models, respectively, entailing that their frequencies are lower. This can be seen by the positive $\Delta f$ at $X_{\rm c}<0.6$ in \cref{fig:dfEvol_M12}.
The frequency differences between the p modes are thus most heavily dependent on the different evolution of the stellar radius in the various prescriptions.

In a practical application, we do not know the mass of the star, unless it comes from model-independent information such as eclipsing double-lined binaries \citep{2010A&ARv..18...67T}. Assuming a precision of 5\% on the mass, the radii change as indicated in \cref{table:radii_masses}.
When comparing the radii listen in \cref{table:radii_Bcep} to those in \cref{table:radii_masses}, we see that the difference in radius caused by switching from an exponential to a Gumbel profile remains smaller than the difference in radius of a model whose mass deviates by 1\%. The models whose thermal structure in the boundary mixing region is convective and that are in an evolved state with $X_{\rm c}<0.4$ have radii that differ more than that of a model with a mass deviating by 1\%; this implies that, even for the best cases of eclipsing double-lined binaries, the model-independent mass estimates reaching 1\% precision are in general not sufficiently precise to unravel the CBM based on p-mode frequencies. The best binaries with low-order p modes to use for such a test are those near the terminal age main sequence.
All p-mode frequency differences for the models with the four different CBM profiles are smaller than the frequency differences due to the uncertainty on masses at a level of 5\%; this implies that any asteroseismic tuning of the near-core mixing on the basis of p modes 
requires knowledge of the mass of the star to better than this relative precision.

\begin{table}[ht]
\caption{Radii, in units of solar radii, of the 12\msol $\beta$ Cep models at different $X_{\rm c}$.}
\label{table:radii_Bcep}  
\centering             
\begin{tabular}{c | r r r r}    
\hline 
$X_{\rm c}$ & \makecell{Diffusive\\exponential} & \makecell{Diffusive\\Gumbel} & \makecell{Convective\\Gumbel} & \makecell{Extended\\convective\\penetration} \\  
\hline 
0.7 & 4.228 & 4.228 & 4.218 & 4.217\\ 
0.6 & 4.856 & 4.855 & 4.841 & 4.841\\ 
0.5 & 5.484 & 5.481 & 5.460 & 5.463\\ 
0.4 & 6.282 & 6.274 & 6.242 & 6.248\\ 
0.3 & 7.338 & 7.326 & 7.275 & 7.290\\ 
0.2 & 8.855 & 8.834 & 8.758 & 8.791\\ 
0.1 & 11.207 & 11.155 & 11.036 & 11.107\\ 
\hline 
\end{tabular}
\end{table}

\begin{table}[ht]
\caption{Radii, in units of solar radii, of $\beta$ Cep models with a mass of 12\msol, and with masses deviating 1 and 5\%, at different $X_{\rm c}$.}
\label{table:radii_masses}      
\centering                      
\begin{tabular}{c | r r r r r }   
\hline 
$X_{\rm c}$ & 11.4\msol & 11.88\msol & 12\msol & 12.12\msol & 12.6\msol \\  
\hline 
0.7 & 4.103 & 4.203 & 4.228 & 4.251 & 4.348\\ 
0.6 & 4.717 & 4.827 & 4.856 & 4.882 & 4.992\\ 
0.5 & 5.328 & 5.452 & 5.484 & 5.514 & 5.640\\ 
0.4 & 6.096 & 6.241 & 6.282 & 6.317 & 5.640\\ 
0.3 & 7.114 & 7.293 & 7.338 & 7.379 & 7.552\\ 
0.2 & 8.575 & 8.798 & 8.855 & 8.905 & 9.129\\ 
0.1 & 10.813 & 11.129 & 11.207 & 11.291 & 11.592\\ 
\hline                              
\end{tabular}
\end{table}

With the shortest data set of 90 days, the p-mode frequencies can be used to discern between diffusive and convective profiles for a fixed stellar mass and input physics for the stellar models. They even allow us to distinguish between all of the mixing prescriptions for evolved models with $X_{\rm c}=0.4$ or lower, since the difference in radii is large enough and only continues to grow during further evolution.
When looking at data sets of one year or longer, the p modes can be used to distinguish between all four mixing prescriptions in all evolutionary stages, as can be seen by the high percentages in \cref{table:percentDGDE,table:percentDGCG,table:percentCGECP}. 

\tabcolsep=4pt
\begin{table}[ht]
\caption{Relative frequency differences $\lb \frac{f_1-f_2}{f_1} \rb$ for the 12\msol models, for $n_{pg}$ = 1, -1, and -2 at different $X_{\rm c}$.}            
\label{table:relDf}
\centering              
\begin{tabular}{c |r r r| r r r| r r r} 
\hline 
 & \multicolumn{3}{c|} {DE-DG} & \multicolumn{3}{c|} {CG-ECP} & \multicolumn{3}{c} {DG-CG} \\ 
 $X_{\rm c}$ &  1 & -1  & -2 &  1 & -1  & -2 & 1 & -1  & -2 \\
\hline 
0.7 & 0.01 & 0.01 & 0.00 & 0.04 & 0.04 & 0.07 & 0.34 & 0.50 & 0.83 \\
0.6 & 0.04 & 0.02 & 0.09 & 0.00 & 0.12 & 0.16 & 0.42 & 0.95 & 0.25  \\
0.5 & 0.06 & 0.09 & 0.10 & 0.10 & 0.17 & 0.17 & 0.57 & 0.72 & 0.07 \\
0.4 & 0.19 & 0.12 & 0.13 & 0.14 & 0.24 & 0.17 & 0.74 & 0.42 & 0.16 \\
0.3 & 0.25 & 0.19 & 0.12 & 0.31 & 0.30 & 0.28 & 1.14 & 0.04 & 0.11 \\
0.2 & 0.27 & 0.33 & 0.15 & 0.45 & 0.52 & 0.49 & 0.71 & 1.01 & 0.93 \\
0.1 & 0.72 & 0.55 & 0.29 & 1.02 & 0.88 & 0.55 & 1.59 & 1.16 & 0.12 \\
\hline
\end{tabular}
\end{table}

The $\Delta f$ for the low-frequency g modes is much smaller than for the high-frequency p modes. However, their relative frequency differences are not necessarily smaller, as can be seen in \cref{table:relDf}. Except for a few low order modes, all the comparisons yield differences smaller than the Rayleigh limit of a 90-day data set.
Depending on the mode radial order and the evolutionary stage, a one-year-long data set might permit a distinction to be made between diffusive and convective models, but it is often not enough to distinguish the fine details of the shape of the near-core mixing.
Using the limit for a four-year-long data set, we should be able to distinguish between convective and diffusive models, and for most modes between all four of the prescriptions if looking at the more evolved models.
This is in line with results by \citet{2015A&A...580A..27M,2016ApJ...823..130M} applied to two \textit{Kepler} SPBs. Similar to the SPB stars, the frequency differences become larger for more evolved models, as can be seen from \cref{fig:dfEvol_M12} and the percentages in \cref{table:percentDGDE,table:percentDGCG,table:percentCGECP}.

The mode kernels of the low-order g modes show similar behaviour to the high-order g modes in the SPB models, in the sense that they remain most sensitive to the same near-core region.
In contrast, the low-order p modes do not appear to be able to probe the near-core region. Instead, they only provide a means of measuring the size of the star, and their probing power occurs in the stellar envelope rather than in the region of the CBM.

\subsection{Future applications after basic asteroseismic modelling.}
Asteroseismic modelling of rotating stars with a convective core was developed
recently thanks to the detection and identification of g modes in nominal four-year
\textit{Kepler} light curves of F and B stars. A global methodological modelling
scheme was developed in \citet{2018ApJS..237...15A} and applications of it occur
in \citet{2015A&A...580A..27M,2016ApJ...823..130M}, \citet{2017A&A...603A..13K}, \citet{2018MNRAS.478.2243S}, and \citet{2019MNRAS.485.3248M}.

In practice, we first have to solve a 5D parameter estimation problem for the
mass, core mass, age, metallicity, and rotation, and this is best done by taking
into account the interplay between uncertainties of measured frequencies and
those due to limitations of the input physics of the equilibrium models;
cf. Section 4 in \citet{2018ApJS..237...15A}. The modelling can be split up by
first estimating the near-core rotation frequency as in
\citet{2016ApJ...823..130M}, \citet{2016A&A...593A.120V}, \citet{2017MNRAS.465.2294O}, and  \citet{2018A&A...618A..47C}. Subsequent estimation of the stellar mass, age, and metallicity based on the measured
period spacing $\Pi_0$, spectroscopic T$_{eff}$ and $\log g$, and Gaia
luminosity offers high precision on the global properties of the star
\citep[cf.][Pedersen et al., in preparation]{2019MNRAS.485.3248M},
but current ensembles are still far too limited to reach mass precisions of
$\sim 1\%$. In order to achieve that with the methods outlined in \citet{2019MNRAS.485.3248M} and Pedersen et al. (in preparation), we need to increase the samples to several hundreds of stars and express that they must adhere to the same theory of stellar evolution. The {\it Kepler\/} database has the potential to reach such ensembles for pulsating AF-type stars, but not for B stars.  

Once the global stellar parameters, such as mass, age, core mass, and
metallicity have been estimated, we can deduce the thermal structure near the
convective core, relying on our results in this section, as well as the
efficiency of the mixing in that region and further out in the radiative
envelope, the latter following \citet{2018A&A...614A.128P}.
These steps thus allow us to assess the efficiency of the mixing and temperature
gradient beyond the convective core from a modelling approach, provided that the
star reveals the appropriate non-radial modes to do so and that its mass can be
estimated with a precision of a few percent.
In practical applications, we must also keep in mind that the
  oscillation frequencies depend on the metal mixture and opacities chosen as
  input physics. Changing these leads to global shifts in the frequency values
\citep[Fig.\,3 in][]{2015A&A...580A..27M}. 
Figure 9 of \citet{2018ApJS..237...15A} illustrates the frequency differences caused
by changing opacity tables (OPAL versus OP) and chemical mixtures (solar versus
OB stars in the solar neighbourhood). The change in frequency for a given
radial order is comparable to the differences found in this study. However,
changing the chemical mixture or opacity tables causes 
a general shift in frequencies, whereas changing 
the temperature gradient in the core boundary region may imply a change in 
the mode radial orders experiencing mode trapping. We may thus hope to unravel
these two effects, although this requires further study.

Once D$_{\text{mix}}(r>\rcc)$ and $\n (r>\rcc)$ have been assessed from profiles
as proposed in this work, it is possible to go a step further and use these profiles as a starting
point for an astrophysical interpretation of the entire $\Dmix(r)$ profiles in
terms of the various mixing causes \citep[cf. Fig 6 in][]{2000A&A...361..101M}.

\section{Conclusions}
We have investigated the impact of various convective core boundary mixing
profiles at the bottom of the radiative envelope on pulsation modes of massive
stars, assuming we know the mass and evolutionary stage with high precision
(better than a few \%).  In general, the differences between mode frequencies
due to the diffusive exponential and diffusive Gumbel profile, and between the
convective Gumbel profile and extended convective penetration, are much smaller
than when comparing the diffusive profiles to the convective profiles.

For the models in the $\beta$ Cep regime, p modes have a larger frequency
difference than the g modes when making comparisons between the different mixing
prescriptions because the latter imply different stellar radii and these modes
are very sensitive to the size of the star.  However, their relative frequency
differences are often comparable to those of the g modes, which have excellent
probing power in the layers with core boundary mixing.
Assuming a fixed mass and input physics, a light curve of 90 days is enough to
discern between both the temperature gradients and, for evolved models, the
mixing profiles, since the differences in radii are largest for the most evolved
models.

Using g modes obtained from a light curve of one year or longer, we should in
general be able to tell the difference between models with convective or
diffusive element mixing. However, the differences due to the functional form of
the diffusion coefficient being a Gumbel or exponential profile are often too
small to distinguish. Therefore we did not refine the mixing profile to switch
halfway through the near-core mixing region from Gumbel to exponential, as
discussed in \cref{sect:gumbel_profile}.  Because of their different internal
structure, the models experience a slightly different evolution. It is therefore
no surprise that more evolved models have more deviating frequencies, and hence
are more easily discerned from one another.

Overall, our study illustrates the promise of asteroseismology applied to large
ensembles of stars that reveal well-identified coherent pulsation modes, whose
frequencies can be measured with high precision. It is noteworthy that coherent p and g modes in B stars have lifetimes much longer
than a typical observing run of months to years.  In that case, data from
different observing runs but with the same equipment can often be combined to
improve the frequency resolution \citep[][for a ground-based data set spanning
21 years]{2003Sci...300.1926A}. In general, the combination from various
observing campaigns does not necessarily lead to a better frequency precision
if the gaps span a longer time base than the re-occuring observation strings
or if the duty cycle between the various sets is too different. Whether or not
this is beneficial has to be assessed on a case-by-case basis, as it depends
on the nature of the modes in terms of amplitude and phase stability.

In order to unravel the core boundary mixing and the thermal structure in the
overshoot zone, we must attempt to have asteroseismic masses and radii with
$\sim 1\%$ relative precision. Following the methods in
\citet{2018ApJS..237...15A} and \citet{2019MNRAS.485.3248M}, we estimate that we
need samples of a few hundred stars to achieve this, as
  good coverage of the rotational frequency with respect to the critical rate must be
  achieved, as well as a variety in metallicity and metal mixture.  While
numerous AF-type g-mode pulsators are still buried in the entire nominal
\textit{Kepler} database, this mission only delivered a few tens of B-type
stars. The capacity of the TESS mission to achieve appropriate samples of
intermediate-mass and high-mass stars with suitable pulsation modes is promising
\citep{2019ApJ...872L...9P}.  In this framework, new coupling such as the action
of rotation on the convective core boundary mixing
\citep[e.g.][]{2004ApJ...601..512B,2016ApJ...829...92A,2019ApJ...874...83A} may
be explored. For the stars rotating faster than half their
critical rate, it will also be worthwhile to consider any latitudinal
dependence of the mixing and temperature gradient in the convective core
boundary region in future applications
\citep[cf.][]{1998ApJ...499..340D,2000ApJ...543..395D,2013A&A...552A..35E}.

\begin{acknowledgements}
  The authors are grateful to Bill Paxton and Rich Townsend, and their team of
  \texttt{MESA} and \texttt{GYRE} developers for their efforts
  and for releasing their
  software publicly; this study would
  not have been possible without their codes.
The research leading to these results has received
  funding from the European Research Council (ERC) under the European Union's
  Horizon 2020 research and innovation programme (grant agreements no. 670519:
  MAMSIE and grant agreement no. 647383: SPIRE). S.M and K.A acknowledge support
  from the CNES PLATO grant at CEA/DAp. We thank the referee for the constructive report which has allowed us to improve the paper.

\end{acknowledgements}


\bibliographystyle{aa}

\begin{appendix}
\section{MESA and GYRE Inlists} \label{appendix:inlist} 

Example MESA and GYRE inlists used for this work are
available from the MESA Inlists section of the MESA Marketplace: 
\url{cococubed.asu.edu/mesa_market/inlists.html.}
\end{appendix}

\end{document}